\documentclass[journal]{IEEEtran}

\usepackage{fancyhdr}
\usepackage{amsmath,amssymb}
\usepackage{amsfonts}
\usepackage[dvips]{graphicx}
\usepackage{dsfont}
\usepackage{comment}
\usepackage{algorithmic}
\usepackage{comment}
\usepackage{caption}
\usepackage{comment}
\usepackage[noabbrev]{cleveref}
\usepackage{balance}
\usepackage{algorithm}
\usepackage{float}
\usepackage{color}
\usepackage{pgfplots}         
\pgfplotsset{compat=1.16}

\usepackage{enumerate}
\usepackage{graphics}
\usepackage{subcaption}
\usepackage{cite}
\usepackage{mathrsfs}
\usepackage{stfloats}

\usepackage{tikz}
\usetikzlibrary{positioning}
\usetikzlibrary{shapes.geometric}
\usetikzlibrary{shapes.misc}
\usepackage{mathdots}
\usepackage{yhmath}
\usepackage{cancel}
\usepackage{color}
\usepackage{siunitx}
\usepackage{array}
\usepackage{multirow}
\usepackage{textcomp}
\usepackage{gensymb}
\usepackage{tabularx}
\usepackage{booktabs}
\usetikzlibrary{fadings}
\usetikzlibrary{patterns}
\usetikzlibrary{shadows.blur}
\usepackage{graphicx}
\usetikzlibrary{decorations.pathreplacing,calligraphy}
\usetikzlibrary{arrows.meta, decorations.markings}
\usepackage{tikz}
\usepackage{tkz-euclide}
\usepackage{pgfplots}

\usepackage{enumitem}

\newtheorem{definition}{Definition}

\newtheorem{proposition}{Proposition}
\newtheorem{remark}{Remark}

\newcommand{\pn}{\sigma_{\phi}}

\newcommand{\Asquare}{A^2(M,\Gamma)}
\usepackage{amsmath,amssymb}
\usepackage{amsfonts}
\usepackage[dvips]{graphicx}
\usepackage{dsfont}
\usepackage{algorithmic}
\usepackage{balance}
\usepackage{algorithm}
\usepackage{mathtools}
\usepackage{comment}

\usepackage{color}
\usepackage{pgfplots}           
\pgfplotsset{compat=1.16}
\usepackage{caption}

\usepackage{enumerate}
\usepackage{graphics}
\usepackage{cite}
\usepackage{mathrsfs}
\usepackage{subcaption}
\usepackage{stfloats}
\usepackage{xfrac}

\usepackage{tikz}
\usepackage{mathdots}
\usepackage{yhmath}
\usepackage{cancel}
\usepackage{color}
\usepackage{siunitx}
\usepackage{array}
\usepackage{multirow}
\usepackage{textcomp}
\usepackage{gensymb}
\usepackage{tabularx}
\usepackage{booktabs}
\usetikzlibrary{fadings}
\usetikzlibrary{patterns}
\usetikzlibrary{shadows.blur}
\usepackage{graphicx}
\DeclarePairedDelimiter\ceil{\lceil}{\rceil}
\DeclarePairedDelimiter\floor{\lfloor}{\rfloor}
\usepackage{mathtools, nccmath}

\begin{document}

\title{On the Design of Super Constellations}
\author{
		Thrassos K. Oikonomou,~\IEEEmembership{Graduate Student Member,~IEEE,}
            Dimitrios Tyrovolas,~\IEEEmembership{Member,~IEEE,}\\
		Sotiris A. Tegos,~\IEEEmembership{Senior Member,~IEEE,}
		Panagiotis D. Diamantoulakis,~\IEEEmembership{Senior Member,~IEEE,}\\
        Panagiotis Sarigiannidis,~\IEEEmembership{Member,~IEEE,} 
        and George K. Karagiannidis,~\IEEEmembership{Fellow,~IEEE}
\thanks{T. K. Oikonomou, D. Tyrovolas, and P. D. Diamantoulakis are with the Department of Electrical and Computer Engineering, Aristotle University of Thessaloniki, 54124 Thessaloniki, Greece (e-mails: \{toikonom, tyrovolas, padiaman\}@auth.gr).}
\thanks{S. A. Tegos is with the Department of Electrical and Computer Engineering, Aristotle University of Thessaloniki, 54124 Thessaloniki, Greece and also with the Department of Electrical and Computer Engineering, University of Western Macedonia, 50100, Kozani, Greece (e-mail: sotiristegos@ieee.org).}
\thanks{P. G. Sarigiannidis is with the Department of Electrical and Computer Engineering, University of Western Macedonia, 50100, Kozani, Greece (e-mail: psarigiannidis@uowm.gr).}
\thanks{G. K. Karagiannidis is with the Department of Electrical and Computer Engineering, Aristotle University of Thessaloniki, 54124 Thessaloniki, Greece and also with the Artificial Intelligence \& Cyber Systems Research Center, Lebanese American University (LAU), Lebanon (e-mail: geokarag@auth.gr).}
\thanks{This work was funded from the Smart Networks and Services Joint Undertaking (SNS JU) under European Union's Horizon Europe research and innovation programme (Grant Agreement No. 101096456 - NANCY).}
\thanks{The work of G. K. Karagiannidis was implemented in the framework of H.F.R.I call basic research financing (horizontal support of all sciences) under the National Recovery and Resilience Plan Greece 2.0 funded by the European Union NextGenerationEU (H.F.R.I. Project Number: 15642).}

}
\maketitle

\begin{abstract} 
In the evolving landscape of sixth-generation (6G) wireless networks, which demand ultra-high data rates, this study introduces the concept of super-constellation communications. Also, we present super amplitude phase shift keying (SAPSK), an innovative modulation technique designed to achieve these ultra-high data rate demands. SAPSK is complemented by the generalized polar distance detector (GPD-D), which approximates the optimal maximum likelihood detector in channels with Gaussian phase noise (GPN). By leveraging the decision regions formulated by GPD-D, a tight closed-form approximation for the symbol error probability (SEP) of SAPSK constellations is derived, while a detection algorithm with $\mathcal{O}(1)$ time complexity is developed to ensure fast and efficient SAPSK symbol detection. Finally, the theoretical performance of SAPSK and the efficiency of the proposed $\mathcal{O}(1)$ algorithm are validated by numerical simulations, highlighting both its superiority in terms of SEP compared to various constellations and its practical advantages in terms of fast and accurate symbol detection.
\end{abstract}
\begin{IEEEkeywords}
super constellations, phase noise, ultra-high data rates, detection
scheme, symbol error probability
\end{IEEEkeywords}

\section{Introduction}

Due to the emergence of promising applications such as augmented reality, holographic communication and super-high-definition live streaming, achieving high data rates is crucial for future digital societies \cite{6G, liaskos}. In this direction, research has focused on employing higher-order modulation schemes. Notably, IEEE 802.11 standard included 4096-quadrature amplitude modulation (QAM), aiming to achieve higher peak data rates.  Concurrently, the architectural design of 6G networks is being tailored to function as a multi-band framework that provides wider bandwidth and improved signal-to-noise ratio (SNR) at the receiver through novel technologies such as massive multiple-input multiple-output (mMIMO), reconfigurable intelligent surfaces (RIS) and higher frequency bands, e.g., free space optics (FSO) \cite{mmWave, massive_mimo_karag}. These technological advancements pave the way for a transformation in modulation schemes, transcending traditional modulation orders to embrace higher-order constellations that can significantly increase the achievable rates, enabling the transmission of more bits per symbol. 
In essence, to meet the demands of next-generation applications with an emphasis on environmental sustainability, it is imperative to design advanced modulation schemes capable of delivering ultra-high data rates in the most energy-efficient manner.

\subsection{State-of-the-Art}

In recent years, increasing the constellation order has been investigated as a means to increase achievable rates substantially. In particular, recent advances in constellation design have focused on the development and optimization of high-order QAM schemes, with significant progress made in the design of 1024-QAM and 4096-QAM constellations \cite{1k-QAM_fiber_optics,1k-QAM-ofdm, 1k-QAM-low-complexity-demap, 1k-QAM_uhdtv,1k-QAM_bicm,1k-non_uniform_QAM}. For example, in \cite{1k-QAM_fiber_optics}, the authors optimized the shape of a 1024-QAM constellation in a fiber channel to achieve performance close to Shannon capacity. Similarly, in \cite{1k-QAM-ofdm}, the authors investigated a shaped 1024-QAM constellation that significantly increases data rates in orthogonal frequency division multiplexing systems while meeting peak-to-average power ratio and symbol error probability (SEP) requirements. In addition, \cite{1k-QAM-low-complexity-demap} and \cite{1k-QAM_uhdtv} designed non-uniformly shaped 256-QAM and 4096-QAM, respectively, to maximize the mutual information of input signals to achieve increased data rates in broadcasting systems. Finally, both \cite{1k-QAM_bicm} and \cite{1k-non_uniform_QAM} have implemented constellation shaping for 1024-QAM together with advanced coding techniques to increase the capacity of the system.

Regarding the novel technologies that will be incorporated within 6G networks, such as mMIMO systems and high frequency bands, extensive research has been conducted into how these key technologies can enable the use of higher-order constellations to achieve ultra-high data rates. Specifically, in \cite{MIMO-256QAM} and \cite{MIMO-equalizer-high-orders}, efficient and low-complexity mMIMO receivers were proposed when 256-QAM constellations are deployed, while in \cite{MIMO-PSK} the rate capabilities of mMIMO satellite network using the 256-PSK constellation were investigated. In addition, the authors of \cite{mmWave-256QAM} developed an mmWave testbed operating at 220 GHz and using 256-QAM modulation, achieving a significant increase in data rates, while also investigating the practical challenges of the system in detail. Furthermore, in \cite{4k-QAM-ofdm}, the transmission of 4096-QAM OFDM operating at 117 GHz is demonstrated, achieving a substantial increase in data rates compared to the existing OFDM systems. Similarly, in \cite{1k-OFDM-28Ghz}, a 1024-QAM OFDM-based system operating at 28 GHz using 5G mmWave phased array antenna, with the main emphasis being given on the benefits in higher peak data rates when high-order QAM are utilized in OFDM. While high-order, the constellations used in these studies are conventional and not carefully designed for achieving ultra-high data rates in an energy-efficient manner. In the context of energy-efficient higher-order modulations, the authors of \cite{hqam_th} performed a detailed theoretical analysis of the SEP for hexagonal-QAM (HQAM) and emphasized that optimal symbol allocation can significantly improve system performance by achieving higher data rates with lower energy consumption. Similarly, in \cite{FSO-high-order}, different modulation schemes such as 1024-HQAM, 1024-cross QAM (XQAM) and 1024-Star QAM were compared and the superiority of HQAM in terms of energy efficiency in a mixed RF/FSO system was shown. Finally, the authors in \cite{HQAM_ris}, highlighted the reduction of system's energy when HQAM is employed in RIS-assisted systems compared to conventional QAM.

\subsection{Motivation \& Contribution}
In the aforementioned works, high-order constellations were used in response to the increasing demands for ultra-high data rate transmissions. However, the transition to the era of super-constellation communications, where very high-order modulation designs are employed, requires an accurate characterization of their inherent susceptibility to specific system imperfections. One of the main impairments that can lead to significant performance degradation in super constellations is the presence of phase noise (PN) due to local oscillator instabilities and the lack of accurate channel state information \cite{LO, AMC-pn,1kand4k-QAM-ofdm-pn}. This inherent susceptibility to PN distortions can be attributed to the increased number of symbols that are far from the origin in super constellations so that even small PN distortions can lead to symbol detection errors \cite{pn_qam}. Even though some previous works in the literature, such as \cite{Eriksson_opt} and \cite{spiral_const}, attempted to optimize the constellation lattice to minimize the SEP under certain GPN conditions, optimizing very high-order constellations poses significant challenges. The unstructured lattices of these optimized constellations necessitate extensive comparisons to detect the transmitted symbols, particularly in very high-order scenarios. Therefore, it is crucial to develop detection methods that maintain low complexity regardless of the constellation order and perform effectively even under PN conditions. In this direction, the authors of \cite{bicais_pqam} proposed a structured amplitude phase shift keying (APSK) constellation called polar QAM (PQAM), which is characterized by increased PN resistance and low-complexity detection but is particularly energy inefficient due to its suboptimal lattice packing. Considering the challenges posed by these research efforts an urgent need is revealed for a new class of modulation designs that balance PN resistance, low detection complexity, and energy consumption. To be classified as super constellations, these designs must fulfill three critical criteria: i) they must have an order of more than 1024 symbols and be capable of achieving ultra-high data rates, ii) they must have low complexity symbol detection and iii) they must exhibit PN robustness. To the best of the authors' knowledge, there is no work that proposes a novel modulation design that comprehensively fulfills the rules of super constellation communication.

In this paper, we propose for the first time in the existing literature a novel modulation design, namely super APSK (SAPSK), which conforms to the rules of super constellation communication. Specifically, the contribution of our work is as follows.
\begin{itemize}
 \item We define the structure of SAPSK that preserves optimal symbol assignment in terms of energy consumption and PN robustness.
 \item We propose a novel detector, the generalized polar distance detector (GPD-D), which expresses a weighted Euclidean distance in the polar plane and provides an analytical framework for both theoretical analysis and low-complexity detection algorithms in super constellations.
\item Based on the decision regions formulated by GPD-D in the polar plane, we provide a tight closed-form approximation for the SEP of SAPSK constellations under the influence of additive white Gaussian noise (AWGN) and GPN. Also, we develop a detection algorithm with $\mathcal{O}(1)$ time complexity tailored to SAPSK constellations, which is a significant step towards the practical use of SAPSK in super constellation communication.
\item We provide simulation results to evaluate the performance of SAPSK in super-constellation communications. These results demonstrate the substantial superiority of the proposed SAPSK in terms of energy efficiency compared to PQAM and conventional QAM.
\end{itemize}

\subsection{Structure}
The remainder of this paper is organized as follows. The system model and some preliminaries on GPN channels are described in Section II. In Section III,  SAPSK constellation, and the proposed low-complexity detector for super-constellation communications are presented. In Section IV, the performance analysis of SAPSK constellations when affected by AWGN and GPN is provided, and an $O(1)$ time complexity algorithm is developed for SAPSKs. In Section IV, simulation results are provided to verify the derived theoretical results. Finally, Section V concludes the paper.

\section{System model and Preliminaries}
We consider a communication system that transmits a symbol $s\in\mathbb{C}$ that belongs to an $M$-ary constellation denoted by $\mathcal{C}$, with average symbol energy $E_{s}$. Specifically, every constellation $\mathcal{C}$ is composed of $M$ symbols that are appropriately allocated upon distinct energy levels, described as circles of radius $\sqrt{E_q}$, where $E_{q}$ denotes the $q$-th energy level of $\mathcal{C}$. By taking into account an AWGN channel impaired by PN, the received signal $r\in\mathbb{C}$ can be expressed as
\begin{equation}\label{system_model}
    r = \lvert s\rvert e^{j(\phi+\arg\{s\})} + n,
\end{equation}
where $\lvert \cdot \rvert$ and $\arg\{\cdot\}$ denote the amplitude and the phase of a complex number, respectively, $n$ represents a complex Gaussian random variable (RV) with zero-mean and variance $\sigma_{n}^2$, and  $\Bar{\gamma} = E_{s}/N_{0}$ denotes the SNR, where $N_{0} = \sigma_{n}^2$ denotes the AWGN power spectral density. Moreover, $\phi$ expresses the PN caused by imperfect carrier phase recovery, usually modeled as a Tikhonov distributed RV with zero mean and standard deviation $\sigma_{\phi}^2$ \cite{proakis2002communication}. However, in modern communication systems, we focus on PN with $\sigma_{\phi}^2\ll1$, and thus $\phi$ can be accurately described as a GPN with zero mean and variance $\sigma_{\phi}^2$ \cite{akyldiz}. In more detail, in modern communication systems a strong PN scenario is described by $\pn^2 = 10^{-1}$, while a moderate and low PN scenarios are described by $\pn^2 = 10^{-2}$ and $\pn^2=10^{-4}$ \cite{bicais_pn_levels}, respectively.

Typically, in digital communication systems where GPN is neglected, the Euclidean distance detector (EUC-D) is used for symbol detection, which is tailored for the AWGN channel. In more detail, EUC-D decides based on the minimum Euclidean distance from the received symbol $r$, and the estimated symbol can be expressed as
\begin{equation}\label{EUC-D}
    \begin{aligned}
        \hat{s} = \arg \min_{s\in\mathcal{C}}\hspace{4px}\lvert r-s\rvert^2,
    \end{aligned}
\end{equation}
However, EUC-D performs sub-optimal in cases where GPN affects the communication system performance. To this end, it becomes imperative to devise a detection scheme that effectively addresses the impact of GPN. In this direction, by taking into account \eqref{system_model},
a detector tailored for GPN channels, denoted as GAP-D, can be derived as in \cite{Eriksson_metrics}
\begin{equation}\label{optimal_detection_metric}
\small
    \begin{aligned}
    \hat{s} = \arg \min_{s\in\mathcal{C}}\hspace{4px}&\frac{\big(\lvert r \rvert-\lvert s \rvert \big)^2}{\frac{\sigma_{n}^2}{2}} + \frac{\big(\arg\{r\}-\arg\{s\}\big)^2}{\sigma_{\phi}^2 + \frac{\sigma_{n}^2}{2\lvert s \rvert^2}} \\
    &+ \log\left(\sigma_{\phi}^2 + \frac{\sigma_{n}^2}{2\lvert s \rvert^2}\right).
\end{aligned}
\end{equation}
As it can be observed in \eqref{optimal_detection_metric}, the performance of GAP-D is influenced by both $\sigma_{n}^2$ and $\sigma_{\phi}^2$, while it intricately incorporates the amplitude and phase of the received symbol $r$ and the constellation symbols. Finally, it should be highlighted that the incorporation of amplitude and phase in \eqref{optimal_detection_metric} translates into an analytical focus on the symbols' polar coordinates.

\begin{figure}[h!]
	\centering
	\begin{tikzpicture}[scale=0.7]
	\begin{semilogyaxis}[
    width=1\linewidth,
	xlabel = $\frac{E_{s}}{N_{0}}$ (dB),
	ylabel = Symbol Error Probability,
	xmin = 0,
	xmax = 80,
	ymin = 0.00001,
	ymax = 1,
    xtick = {0,10,...,80},
	grid = major,
	legend entries = {{32-QAM (EUC-D)}, {32-QAM (GAP-D)}, {4096-QAM (EUC-D)}, {4096-QAM (GAP-D)}},
	legend cell align = {left},
    legend pos = south east,
    legend style={font=\tiny}
	]
	
	\addplot[
        color = black,
        no marks,
	mark repeat = 2,
	mark size = 3,
        line width = 1pt,
        style = solid
	]
	table {plots/transaction-data/qam/32-QAM-sim-eucd_1.000000e-04_.txt};
        \addplot[
        color = black,
	mark = diamond*,
	mark repeat = 2,
	mark size = 3,
        line width = 1pt
	]
	table {plots/transaction-data/qam/32-QAM-sim-GAPD_1.000000e-04_.txt};
        \addplot[
        color = red,
	no marks,
	mark repeat = 1,
	mark size = 2,
        style=solid,
        line width = 1pt
	]
	table {plots/transaction-data/qam/4096-QAM-sim-EUCD_1.000000e-04_smoothed.txt};
     \addplot[
        color = red,
	mark=square*,
	mark repeat = 2,
	mark size = 2,
        style=solid,
        line width = 1pt
    ]
	table {plots/transaction-data/qam/4096-QAM-sim-GAPD_1.000000e-04_smoothed.txt};
	\end{semilogyaxis}
	\end{tikzpicture}
	\caption{EUC-D vs GAP-D with PN variance $\sigma^2_{\phi} = 10^{-4}$}
	\label{fig:euc_vs_gap}
\end{figure}
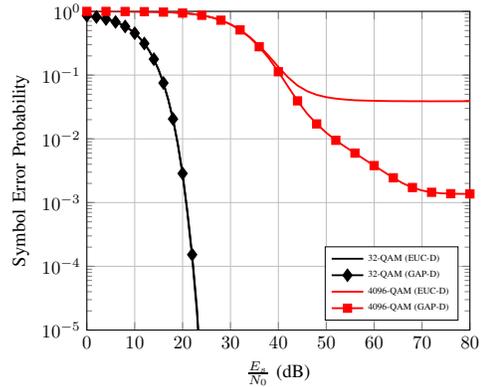

To meet the escalating demands for high data rates in next-generation communication networks, super constellations can be employed, allowing each transmitted symbol to contain more information bits. However, it is imperative to tackle the significant challenges posed by GPN on super constellations while prioritizing the development of energy-efficient designs that minimize the SEP for certain system's energy consumption. Specifically, when the constellation order $M$ is large, i.e., $M \geq 1024$, and EUC-D is applied, even minor GPN leads to significant performance degradation, as shown in Fig. \ref{fig:euc_vs_gap}. For example, when dealing with minor GPN, e.g., $\pn^2 = 10^{-4}$, low-order QAM constellations exhibit satisfactory SEP performance, a phenomenon that is not observed in the case of high-order QAM, signifying that EUC-D can not be used in super-constellation communications. Consequently, in super-constellation communications, it is imperative to consider a detector that mitigates the GPN effect. In this direction, as it can be observed in Fig. \ref{fig:euc_vs_gap}, GAP-D can significantly reduce the SEP of high-order QAM constellations, emphasizing the importance of detectors that effectively minimize the impact of GPN.

Although GAP-D minimizes the effect of GPN in the system, it does not ensure its elimination, leading to the presence of an error floor as shown in Fig. \ref{fig:euc_vs_gap}, which indicates that increasing the SNR will not always translate into SEP improvement. To this end, the employment of GAP-D needs to be combined with an efficient modulation design that can further reduce this error floor to achieve robust super-constellation communications. In more detail, as illustrated in Fig. \ref{fig:QAM-PTQAM-iq-polar}b, when GAP-D is applied for a conventional constellation, e.g., QAM, the system performance is severely affected by an increased probability of erroneous detection, since the symbols in the polar plane are not separated by large distances. Therefore, it is necessary to design constellation structures with symbols appropriately allocated in the polar plane, demonstrating inherent GPN resilience and a reduced error floor when GAP-D is applied. Moreover, given the increased SNR requirements for robust super constellations, it is crucial to allocate symbols in the polar plane to not only ensure resilience against GPN but also to reduce system's energy needs. In conclusion, optimizing symbol allocation in the polar plane is essential for reducing the error floor occurred by GPN, and thus, achieving efficient super-constellation communications. 

\section{A Novel Modulation Design for super constellation Communication}
\subsection{SAPSK}
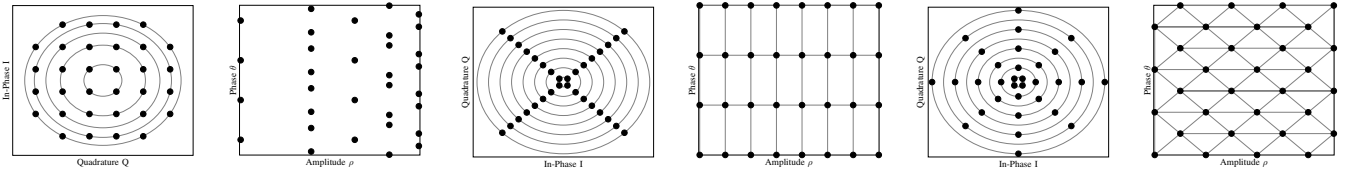
\begin{figure*}
\centering
\begin{minipage}[t]{0.16\textwidth}
\centering
\begin{tikzpicture}[scale=0.35]
\begin{axis}[
        xtick=\empty,
        ytick=\empty,
        xlabel={$\text{Quadrature Q}$},
        ylabel={$\text{In-Phase I}$},
        xlabel style={below},
        ylabel style={above},
        xmin=-1.5,
        xmax=1.5,
        ymin=-1.5,
        ymax=1.5,
    ]
    \addplot [only marks, mark size = 3, black] table {figures/data/QAM_iq.txt};
    \draw[gray] (0,0) circle (0.316);
    \draw[gray] (0,0) circle (0.707);
    \draw[gray] (0,0) circle (0.948);
    \draw[gray] (0,0) circle (1.14);
    \draw[gray] (0,0) circle (1.3);
\end{axis}
\end{tikzpicture}
\label{fig:QAM-iq}
\end{minipage}
\begin{minipage}[t]{0.16\textwidth}
\centering
\begin{tikzpicture}[scale=0.35]
\begin{axis}[
        xtick=\empty,
        ytick=\empty,
        xlabel={$\text{Amplitude $\rho$}$},
        ylabel={$\text{Phase $\theta$}$},
        xlabel style={below},
        ylabel style={above},
        xmin=0.31,
        xmax=1.31,
        ymin=-2.96,
        ymax=2.96
    ]
    \addplot [only marks, mark size = 3,black] table {figures/data/QAM_polar.txt};
\end{axis}
\end{tikzpicture}
\label{fig:QAM-polar}
\end{minipage}
\begin{minipage}[t]{0.16\textwidth}
\centering
\begin{tikzpicture}[scale=0.35]
\begin{axis}[
        xtick=\empty,
        ytick=\empty,
        xlabel={$\text{In-Phase I}$ },
        ylabel={$\text{Quadrature Q}$},
        xlabel style={below},
        ylabel style={above},
        xmin=-1.7,
        xmax=1.7,
        ymin=-1.7,
        ymax=1.7
    ]
    \draw[gray] (0,0) circle (0.1085);
    \draw[gray] (0,0) circle (0.3255);
    \draw[gray] (0,0) circle (0.5425);
    \draw[gray] (0,0) circle (0.7595);
    \draw[gray] (0,0) circle (0.9765);
    \draw[gray] (0,0) circle (1.1935);
    \draw[gray] (0,0) circle (1.4105);
    \draw[gray] (0,0) circle (1.6275);
    \addplot [only marks, mark size = 3, black] table {plots/lattice_pqam/pqam_points_iq_32_8.txt};	
    \end{axis}
    \end{tikzpicture}   
\end{minipage}
\begin{minipage}[t]{0.16\textwidth}
\centering
\begin{tikzpicture}[scale=0.35]
\begin{axis}[
        xtick=\empty,
        ytick=\empty,
        xlabel={$\text{Amplitude $\rho$}$},
        ylabel={$\text{Phase $\theta$}$},
        xlabel style={below},
        ylabel style={above},
        xmin=0.1,
        xmax=1.63,
        ymin=-2.36,
        ymax=2.36
    ]
    \addplot [only marks, mark size = 3, black] table {plots/lattice_pqam/pqam_points_polar_32_8.txt};
    \addplot [no markers, update limits=false, gray] table{plots/lattice_pqam/pqam_points_polar_32_8_lattice.txt};
\end{axis}
\end{tikzpicture}
\end{minipage}
\begin{minipage}[t]{0.16\textwidth}
\centering
\begin{tikzpicture}[scale=0.35]
\begin{axis}[
        xtick=\empty,
        ytick=\empty,
        xlabel={$\text{In-Phase I}$ },
        ylabel={$\text{Quadrature Q}$},
        xlabel style={below},
        ylabel style={above},
        xmin=-1.7,
        xmax=1.7,
        ymin=-1.7,
        ymax=1.7
    ]
    \draw[gray] (0,0) circle (0.1085);
    \draw[gray] (0,0) circle (0.3255);
    \draw[gray] (0,0) circle (0.5425);
    \draw[gray] (0,0) circle (0.7595);
    \draw[gray] (0,0) circle (0.9765);
    \draw[gray] (0,0) circle (1.1935);
    \draw[gray] (0,0) circle (1.4105);
    \draw[gray] (0,0) circle (1.6275);
    \addplot [only marks, mark size = 3, black] table {figures/data/points_iq_32_8.txt};	
    \end{axis}
    \end{tikzpicture}   
\end{minipage}
\begin{minipage}[t]{0.16\textwidth}
\centering
\begin{tikzpicture}[scale=0.35]
\begin{axis}[
        xtick=\empty,
        ytick=\empty,
        xlabel={$\text{Amplitude $\rho$}$},
        ylabel={$\text{Phase $\theta$}$},
        xlabel style={below},
        ylabel style={above},
        xmin=0.1,
        xmax=1.63,
        ymin=-2.36,
        ymax=3.15
    ]
    \addplot [only marks, mark size = 3, black] table {plots/lattice_ptqam/ptqam_points_polar_32_8.txt};
    \addplot [no markers, update limits=false, gray] table{plots/lattice_ptqam/ptqam_points_polar_32_8_lattice.txt};
\end{axis}
\end{tikzpicture}
\end{minipage}
\caption{$32$-QAM in a) IQ plane, b) Polar plane, $32$-PQAM($8$) in c) IQ plane, d) Polar plane and SAPSK($32$,$8$) in e) IQ plane, f) Polar plane}
\label{fig:QAM-PTQAM-iq-polar}
\end{figure*}
Building upon the need to develop modulation schemes for robust super-constellation communications, it becomes crucial to examine constellation lattices with appropriate formation in the polar plane. In this direction, APSK constellation structures have appeared to offer enhanced performance in the presence of GPN, in contrast to other well-established constellations like QAM \cite{bicais_pqam}. Specifically, APSK constellations are allocated on energy levels, with neighboring symbols on the same ring maintaining a specific angular distance between them. Additionally, this spatial arrangement allows for the dynamic adjustment of the angular distances and the energy levels of the constellations to compensate for the effect of both AWGN and GPN and, also, improve the energy consumption of the constellation. In this context, a specific APSK formation shown in Fig. \ref{fig:QAM-PTQAM-iq-polar}c and referred to as PQAM, has been proposed in \cite{bicais_pqam}, which adopts a rectangular lattice in the polar plane, as illustrated in Fig. \ref{fig:QAM-PTQAM-iq-polar}. However, the energy efficiency of the constellation can be further improved by compactly allocating the symbols in the polar lattice. In this direction, by leveraging the flexibility of APSK in symbol allocation and the increased energy efficiency demonstrated by triangular QAM (TQAM) constellations in AWGN channels, a novel modulation design featuring a triangular structure in the polar domain, denoted as SAPSK can be designed for robust super-constellation communications. Specifically, as illustrated in Fig. \ref{fig:QAM-PTQAM-iq-polar}e, an $M$-ary SAPSK consists of $M$ symbols arranged on $\Gamma$ concentric circles of radius $\rho_q$ which is given as
\begin{equation} \label{radius}
    \begin{aligned}
        \rho_q = \frac{\delta_{\rho}}{2}(2q-1),
    \end{aligned}
\end{equation}
where $q \in \{1, 2,\dotsc, \Gamma\}$ and $\rho_1$ corresponds to the radius of the lowest energy level, while
$\delta_{\rho}$ is the radial distance between two adjacent concentric circles which is equal to
\begin{equation}\label{dr}
    \begin{aligned}
        \delta_{\rho} = \sqrt{\frac{ 12 E_{s}}{4\Gamma^2-1}}.
    \end{aligned}
\end{equation}
Therefore, each symbol on the $q$-th concentric circle of the SAPSK has energy $E_{q}$ which can be calculated 
\begin{equation}\label{Eq_sxesi}
    \begin{aligned}
        E_q = \frac{3E_s\left(2q-1\right)^2}{4\Gamma^2-1}
    \end{aligned}
\end{equation}
Furthermore, each concentric circle of a PTQAM constellation contains $\frac{M}{\Gamma}\in\mathbb{N}^{+}$ symbols equally spaced by angle $\delta_{\theta}$ that can be expressed as
\begin{equation}\label{dth}
    \begin{aligned}
        \delta_{\theta} = \frac{2\pi\Gamma}{M}
    \end{aligned}
\end{equation} 
and the argument of $p$-th symbol located at the $q$-th energy level is given by 
\begin{equation}\label{phase_offset}
    \begin{aligned}
        \arg\{s_{pq}\} = 
        \frac{\delta_{\theta}}{2}(2p-1)\left(1+\mathrm{mod}(q,2)\frac{\delta_{\theta}}{2}\right),
    \end{aligned}
\end{equation}
where $p \in \{1,2,\dotsc, \frac{M}{\Gamma} \}$, and $\mathrm{mod}\left(\cdot,\cdot \right)$ is the modulo operation\color{black}. Consequently, an SAPSK$\left(M, \Gamma\right)$ constellation is defined by $M$ and $\Gamma$ and can be described as
\begin{equation}\label{PHQAM}
\small
    \begin{aligned}
        \mathcal{C}' = \bigg\{\frac{\delta_{\rho}}{2}(2q-1) \exp&\left(j \frac{\delta_{\theta}}{2}(2p-1)\left(1+\mathrm{mod}(q,2)\frac{\delta_{\theta}}{2}\right)\right)\\
        &\bigg| 1\leq q\leq\Gamma, 1\leq p \leq \frac{M}{\Gamma}\bigg\}.
    \end{aligned}
\end{equation}
Finally, by applying a polar transformation in the symbols of $\mathcal{C}'$ a triangular lattice is obtained in the polar plane, as illustrated in Fig. \ref{fig:QAM-PTQAM-iq-polar}f. To this end, the definition of SAPSK$\left(M, \Gamma\right)$ constellation based on its polar lattice is as follows:
\begin{definition} 
\textit{SAPSK$\left(M, \Gamma\right)$ is defined as a constellation in the polar plane with $M$ symbols equally distributed on $\Gamma$ columns that are spaced by $\delta_{\rho}$, and between adjacent symbols isosceles triangles are formed.}
\end{definition} 

As it can be observed in Fig. \ref{fig:QAM-PTQAM-iq-polar}d and Fig. \ref{fig:QAM-PTQAM-iq-polar}f, depicting the polar lattices of PQAM and SAPSK, respectively, both PQAM and SAPSK maintain an equal distance between symbols in the same column, ensuring GPN resilience. However, it is evident from the figures that SAPSK exhibits enhanced robustness compared to PQAM in horizontal movements, as its lattice doubles the distance between symbols in the horizontal direction induced by the AWGN. In summary, the SAPSK lattice allows for a more compact symbol allocation in the polar plane compared to PQAM, achieving increased robustness against erroneous symbol detection induced by both AWGN and GPN, while preserving the same average constellation energy.

\subsection{Low-complexity Detector Design for super-constellation communications in GPN}
We need detection algorithms with low time complexity to make the implementation of super constellations in practical systems feasible. To achieve this, we can capitalize on the geometric characteristics of the constellation decision regions to avoid comparing all the symbols with the received one to detect which symbol was transmitted. However, considering the sensitivity of super constellations to GPN and the GAP-D detector, the detection process in super-constellation communications must account for the decision regions formulated in the polar plane, by utilizing an appropriate detector tailored for GPN mitigation. 

Given the high-order of super constellations, developing detection schemes that maintain low complexity and operate effectively under GPN conditions is crucial. In this direction, while the detection metric offered by GAP-D is applicable for optimal symbol detection in GPN scenarios, its optimality is countered by the high time complexity and the complex decision regions it introduces. Specifically, it becomes apparent in \eqref{optimal_detection_metric} that GAP-D requires $M$ calculations to determine $\hat{s}$, resulting in a complexity of $\mathcal{O}(M)$, rendering it unsuitable for super constellations. Additionally, the dependence of GAP-D on the magnitude of every constellation symbol $s$ and the logarithmic factor signify that GAP-D defines decision regions that cannot be utilized to produce low-complexity detection algorithms. In more detail, the decision regions formulated by GAP-D are different for each symbol, hence the composed decision region lattice has unstructured and complex geometry.
Therefore, it is imperative to develop a detector that performs nearly optimal, while enabling the production of decision regions with common geometry in the polar plane, e.g., polygons. To this end, in the following proposition, we introduce GPD-D, which tightly approximates the performance of GAP-D and expresses a weighted Euclidean distance in the polar plane, enabling the construction of polygonic decision regions.
\begin{proposition}
The transmitted symbol $s\in\mathcal{C}$ affected by both AWGN and GPN can be reliably estimated through GPD-D which is a weighted polar Euclidean distance detector, and is expressed as
\begin{equation}\label{proposed_detection_metric}
\small
    \begin{aligned}
    \hat{s} = 
            \arg \min_{s\in\mathcal{C}}\hspace{4px}\frac{2\big(\lvert r \rvert-\lvert s \rvert \big)^2}{\sigma_{n}^2} + \frac{\big(\arg\{r\}-\arg\{s\}\big)^2}{\sigma_{\phi}^2 + \frac{\sigma_{n}^2}{2\lvert r \rvert^2}},
\end{aligned}
\end{equation}
\end{proposition}
\begin{IEEEproof}
The GAP-D metric results in symbol-dependent structures unsuitable for developing time-efficient detection algorithms. In this direction, we can substitute the factor $|s|^2$ in GAP-D by a constant term to produce constant symbol-independent decision regions. In this context, it is imperative to understand the effects of the variables on the term $\sigma_{\phi}^2 + \frac{\sigma_{n}^2}{2\lvert s \rvert^2}$, enabling the substitution of $|s|^2$ with a constellation-dependent constant parameter. First, the operating SNR in super-constellation communications must be high to achieve satisfactory performance. Moreover, in the context of super constellation designs, symbols must be allocated across numerous energy levels to prevent errors arising from phase rotation due to GPN with the same energy level, while keeping the energy levels as close as possible to the origin to confine energy requirements.
As a result, the SNR value of two consecutive energy levels can be approximately considered equal,
leading to the following approximation 
\begin{equation}\label{dE}
    \begin{aligned} 
    \frac{\sigma_{n}^2}{E_{q}} \approx \frac{\sigma_{n}^2}{E_{q+1}}.
    \end{aligned}
\end{equation}
Furthermore, using \eqref{dE}, the following approximation can be derived
\begin{equation} \label{approx_problematic_term}
    \begin{aligned} 
        \sigma_{\phi}^2 + \frac{\sigma_{n}^2}{2E_{q}} \approx \sigma_{\phi}^2 + \frac{\sigma_{n}^2}{2E_{q+1}}.
    \end{aligned}
\end{equation}
Consequently, by considering that the energy levels $E_{q}$ and $E_{q+1}$ enclose the received symbol $r$, i.e., $|r|^2 \in [E_{q}, E_{q+1}]$, then for every $s\in\mathcal{P}$, where $\mathcal{P}=\big\{s\in\mathcal{C}: \lvert s\rvert = \sqrt{E_{q}} \lor \lvert s\rvert = \sqrt{E_{q+1}} \big\}$, the following approximation holds

\begin{equation} 
    \begin{aligned} \label{approx_first_branch}
        \sigma_{\phi}^2 + \frac{\sigma_{n}^2}{2\lvert s \rvert^2} \approx \sigma_{\phi}^2 + \frac{\sigma_{n}^2}{2\lvert r \rvert^2}, \hspace{1em}\forall s \in \mathcal{P},
    \end{aligned}
\end{equation}
By substituting \eqref{approx_first_branch} in \eqref{optimal_detection_metric} and omitting the logarithmic term since its argument becomes symbol-independent, \eqref{proposed_detection_metric} can be obtained,  
which concludes the proof.
\end{IEEEproof}

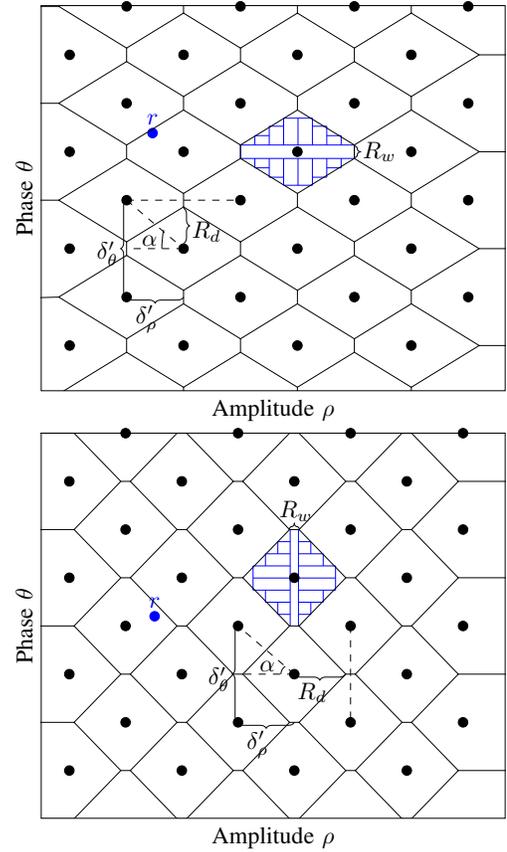
\begin{figure}[t]
\centering
\begin{minipage}[t]{0.5\textwidth}
\centering
\begin{tikzpicture}[scale=0.9]
\begin{axis}[
    xtick=\empty,
    ytick=\empty,
    xlabel={$\text{Amplitude $\rho$}$},
    ylabel={$\text{Phase $\theta$}$},
    xlabel style={below},
    ylabel style={above},
    xmin=0,
    xmax=2.5,
    ymin=-1.42,
    ymax=1.45,
]
\draw[black, dashed]  (4.6017899e-01, -1.1763245e-16)-- (7.6696499e-01, -3.6104166e-01);
\draw[black, dashed]  (2.7,-4.4340237e+00)-- (5.2062690e+00, -4.4340237e+00);
\draw[decorate,
    decoration = {brace}]  (7.6696499e-01, -0.6008332e-01)-- (7.6696499e-01, -3.2208332e-01) node [midway, right, pos=0.67] {$R_{d}$};
\draw[decorate,
    decoration = {brace,mirror}]  (1.6873230e+00, 3.1604166e-01)-- (1.6873230e+00, 4.0504166e-01) node [midway, right] {$R_{w}$};
\draw (6.5e-01, -3.6104166e-01) arc (180:172:1); 
\node[anchor=east] at (6.7e-01, -3.0104166e-01) {$\alpha$};
\draw[black,dashed]  (7.6696499e-01, -3.6104166e-01)-- (4.6017899e-01, -3.6104166e-01);
\draw[decorate,
    decoration = {brace,mirror}]  (4.6017899e-01, -1.1763245e-16)-- (4.6017899e-01, -7.2208332e-01) node [midway, left, pos=0.55] {$\delta_{\theta}'$};
\draw[decorate,
    decoration = {brace}]  (7.6696499e-01, -7.2208332e-01)-- (4.6017899e-01, -7.2208332e-01) node [midway, below, pos=0.65] {$\delta_{\rho}'$};
\draw[black, dashed]  (5.2062690e+00, -2.2170119e+00)-- (5.2062690e+00, -6.6510356e+00);
\draw[black, dashed]  (4.6017899e-01, -1.1763245e-16)-- (10.6696499e-01, -1.1763245e-16);

\draw[blue]  (1.0737510e+00, 3.6104166e-01 + 0.05)-- (1.0737510e+00, 3.6104166e-01 - 0.05);
\draw[blue]  (1.6873230e+00, 3.6104166e-01 + 0.05)-- (1.6873230e+00, 3.6104166e-01 - 0.05);
\draw[blue]  (1.0737510e+00, 3.6104166e-01 + 0.05)-- (1.6873230e+00, 3.6104166e-01 + 0.05);
\draw[blue]  (1.0737510e+00, 3.6104166e-01 - 0.05)-- (1.6873230e+00, 3.6104166e-01 - 0.05);
\draw[blue]  (1.0737510e+00+0.0775, 3.6104166e-01 + 0.05)-- (1.0737510e+00+0.0775, 3.6104166e-01 + 0.12);
\draw[blue]  (1.0737510e+00+0.0775, 3.6104166e-01 + 0.12)-- (1.0737510e+00+2*0.0775, 3.6104166e-01 + 0.12);
\draw[blue]  (1.0737510e+00+2*0.0775, 3.6104166e-01 + 0.05)-- (1.0737510e+00+2*0.0775, 3.6104166e-01 + 0.19);
\draw[blue]  (1.0737510e+00+2*0.0775, 3.6104166e-01 + 0.19)-- (1.0737510e+00+3*0.0775, 3.6104166e-01 + 0.19);
\draw[blue]  (1.0737510e+00+3*0.0775, 3.6104166e-01 + 0.05)-- (1.0737510e+00+3*0.0775, 3.6104166e-01 + 0.25);
\draw[blue]  (1.0737510e+00+3*0.0775, 3.6104166e-01 + 0.25)-- (1.0737510e+00+4*0.0775, 3.6104166e-01 + 0.25);

\draw[blue]  (1.0737510e+00+4*0.0775, 3.6104166e-01 + 0.05)-- (1.0737510e+00+4*0.0775, 3.6104166e-01 + 0.25);
\draw[blue]  (1.0737510e+00+4*0.0775, 3.6104166e-01 + 0.25)-- (1.0737510e+00+5*0.0775, 3.6104166e-01 + 0.25);
\draw[blue]  (1.0737510e+00+5*0.0775, 3.6104166e-01 + 0.25)-- (1.0737510e+00+5*0.0775, 3.6104166e-01 + 0.05);
\draw[blue]  (1.0737510e+00+5*0.0775, 3.6104166e-01 + 0.19)-- (1.0737510e+00+6*0.0775, 3.6104166e-01 + 0.19);
\draw[blue]  (1.0737510e+00+6*0.0775, 3.6104166e-01 + 0.19)-- (1.0737510e+00+6*0.0775, 3.6104166e-01 + 0.05);
\draw[blue]  (1.0737510e+00+6*0.0775, 3.6104166e-01 + 0.12)-- (1.0737510e+00+7*0.0775, 3.6104166e-01 + 0.12);
\draw[blue]  (1.0737510e+00+7*0.0775, 3.6104166e-01 + 0.12)-- (1.0737510e+00+7*0.0775, 3.6104166e-01 + 0.05);
\draw[blue]  (1.0737510e+00+0.0775, 3.6104166e-01 - 0.05)-- (1.0737510e+00+0.0775, 3.6104166e-01 - 0.12);
\draw[blue]  (1.0737510e+00+0.0775, 3.6104166e-01 - 0.12)-- (1.0737510e+00+2*0.0775, 3.6104166e-01 - 0.12);
\draw[blue]  (1.0737510e+00+2*0.0775, 3.6104166e-01 - 0.05)-- (1.0737510e+00+2*0.0775, 3.6104166e-01 - 0.19);
\draw[blue]  (1.0737510e+00+2*0.0775, 3.6104166e-01 - 0.19)-- (1.0737510e+00+3*0.0775, 3.6104166e-01 - 0.19);
\draw[blue]  (1.0737510e+00+3*0.0775, 3.6104166e-01 - 0.05)-- (1.0737510e+00+3*0.0775, 3.6104166e-01 - 0.25);
\draw[blue]  (1.0737510e+00+3*0.0775, 3.6104166e-01 - 0.25)-- (1.0737510e+00+4*0.0775, 3.6104166e-01 - 0.25);

\draw[blue]  (1.0737510e+00+4*0.0775, 3.6104166e-01 - 0.05)-- (1.0737510e+00+4*0.0775, 3.6104166e-01 - 0.25);
\draw[blue]  (1.0737510e+00+4*0.0775, 3.6104166e-01 - 0.25)-- (1.0737510e+00+5*0.0775, 3.6104166e-01 - 0.25);
\draw[blue]  (1.0737510e+00+5*0.0775, 3.6104166e-01 - 0.25)-- (1.0737510e+00+5*0.0775, 3.6104166e-01 - 0.05);
\draw[blue]  (1.0737510e+00+5*0.0775, 3.6104166e-01 - 0.19)-- (1.0737510e+00+6*0.0775, 3.6104166e-01 - 0.19);
\draw[blue]  (1.0737510e+00+6*0.0775, 3.6104166e-01 - 0.19)-- (1.0737510e+00+6*0.0775, 3.6104166e-01 - 0.05);
\draw[blue]  (1.0737510e+00+6*0.0775, 3.6104166e-01 - 0.12)-- (1.0737510e+00+7*0.0775, 3.6104166e-01 - 0.12);
\draw[blue]  (1.0737510e+00+7*0.0775, 3.6104166e-01 - 0.12)-- (1.0737510e+00+7*0.0775, 3.6104166e-01 - 0.05);

\filldraw[blue] (0.6,0.5) circle (2pt) node[anchor=south] {$r$};
\addplot [only marks, mark size = 2,black] table {plots/points/ptqam_points_polar_32_8_snr_0__zone_2_.txt};
\addplot [no markers, update limits=false] table{plots/voronoi/ptqam_voronoi_polar_32_8_snr_0__zone_2_.txt};

\end{axis}

\end{tikzpicture}	
\end{minipage}
\begin{minipage}[t]{0.5\textwidth}
\centering
\begin{tikzpicture}[scale=0.9]
\begin{axis}[
    xtick=\empty,
    ytick=\empty,
    xlabel={$\text{Amplitude $\rho$}$},
    ylabel={$\text{Phase $\theta$}$},
    xlabel style={below},
    ylabel style={above},
    xmin=0,
    xmax=45,
    ymin=-19.85,
    ymax=19.9,
]
\draw[black, dashed]  (2.4549805e+01, -4.9739659e+00)-- (1.9094293e+01, 0);
\draw[black, dashed]  (2.4549805e+01, -4.9739659e+00)-- (1.9094293e+01, -4.9739659e+00);
\draw (2.3249805e+01, -4.979659e+00) arc (180:125:1); 
\node[anchor=east] at (23.5, -4.2e+00) {$\alpha$};
\draw[decorate,
    decoration = {brace,mirror}]  (1.9094293e+01, 0) -- (1.9094293e+01, -9.9479319e+00) node [midway, left, pos=0.55] {$\delta_{\theta}'$};
\draw[decorate,
    decoration = {brace,mirror}]  (1.9094293e+01, -9.9479319e+00) -- (2.4549805e+01, -9.9479319e+00) node [midway, below, pos=0.35] {$\delta_{\rho}'$};
\draw[black, dashed]  (3.0005317e+01, 0) -- (3.0005317e+01, -9.9479319e+00);
\draw[decorate,
    decoration = {brace}]  (2.95e+01, -4.9739659e+00) -- (2.4549805e+01, -4.9739659e+00) node [midway, below, pos=0.65] {$R_{d}$};
\draw[decorate,
decoration = {brace}]  (2.4049805e+01, 9.9479319e+00) -- (2.5049805e+01 , 9.9479319e+00) node [midway, above, pos=0.65] {$R_{w}$};

\draw[blue]  (2.4549805e+01-0.4, 0)-- (2.4549805e+01-0.4, 9.9479319e+00);
\draw[blue]  (2.4549805e+01+0.4, 0)-- (2.4549805e+01+0.4, 9.9479319e+00);
\draw[blue]  (2.4549805e+01 + 0.4, 9.9479319e+00 - 1.25)-- (2.4549805e+01 + 1.6, 9.9479319e+00 - 1.25);
\draw[blue]  (2.4549805e+01 + 1.6, 9.9479319e+00 - 1.25)-- (2.4549805e+01 + 1.6, 9.9479319e+00 - 2*1.25);
\draw[blue]  (2.4549805e+01 + 0.4, 9.9479319e+00 - 2*1.25)-- (2.4549805e+01 + 2.8, 9.9479319e+00 - 2*1.25);
\draw[blue]  (2.4549805e+01 + 2.8, 9.9479319e+00 - 2*1.25)-- (2.4549805e+01 + 2.8, 9.9479319e+00 - 3*1.25);
\draw[blue]  (2.4549805e+01 + 0.4, 9.9479319e+00 - 3*1.25)-- (2.4549805e+01 + 4, 9.9479319e+00 - 3*1.25);
\draw[blue]  (2.4549805e+01 + 4, 9.9479319e+00 - 3*1.25)-- (2.4549805e+01 + 4, 9.9479319e+00 - 4*1.25);
\draw[blue]  (2.4549805e+01 + 0.4, 9.9479319e+00 - 4*1.25)-- (2.4549805e+01 + 4, 9.9479319e+00 - 4*1.25);

\draw[blue]  (2.4549805e+01 + 4, 9.9479319e+00 - 4*1.25)-- (2.4549805e+01 + 4, 9.9479319e+00 - 5*1.25);
\draw[blue]  (2.4549805e+01 + 0.4, 9.9479319e+00 - 5*1.25)-- (2.4549805e+01 + 4, 9.9479319e+00 - 5*1.25);
\draw[blue]  (2.4549805e+01 + 2.8, 9.9479319e+00 - 5*1.25)-- (2.4549805e+01 + 2.8, 9.9479319e+00 - 6*1.25);
\draw[blue]  (2.4549805e+01 + 0.4, 9.9479319e+00 - 6*1.25)-- (2.4549805e+01 + 2.8, 9.9479319e+00 - 6*1.25);
\draw[blue]  (2.4549805e+01 + 1.6, 9.9479319e+00 - 6*1.25)-- (2.4549805e+01 + 1.6, 9.9479319e+00 - 7*1.25);
\draw[blue]  (2.4549805e+01 + 0.4, 9.9479319e+00 - 7*1.25)-- (2.4549805e+01 + 1.6, 9.9479319e+00 - 7*1.25);

\draw[blue]  (2.4549805e+01 - 0.4, 9.9479319e+00 - 1.25)-- (2.4549805e+01 - 1.6, 9.9479319e+00 - 1.25);
\draw[blue]  (2.4549805e+01 - 1.6, 9.9479319e+00 - 1.25)-- (2.4549805e+01 - 1.6, 9.9479319e+00 - 2*1.25);
\draw[blue]  (2.4549805e+01 - 0.4, 9.9479319e+00 - 2*1.25)-- (2.4549805e+01 - 2.8, 9.9479319e+00 - 2*1.25);
\draw[blue]  (2.4549805e+01 - 2.8, 9.9479319e+00 - 2*1.25)-- (2.4549805e+01 - 2.8, 9.9479319e+00 - 3*1.25);
\draw[blue]  (2.4549805e+01 - 0.4, 9.9479319e+00 - 3*1.25)-- (2.4549805e+01 - 4, 9.9479319e+00 - 3*1.25);
\draw[blue]  (2.4549805e+01 - 4, 9.9479319e+00 - 3*1.25)-- (2.4549805e+01 - 4, 9.9479319e+00 - 4*1.25);
\draw[blue]  (2.4549805e+01 - 0.4, 9.9479319e+00 - 4*1.25)-- (2.4549805e+01 - 4, 9.9479319e+00 - 4*1.25);

\draw[blue]  (2.4549805e+01 - 4, 9.9479319e+00 - 4*1.25)-- (2.4549805e+01 - 4, 9.9479319e+00 - 5*1.25);
\draw[blue]  (2.4549805e+01 - 0.4, 9.9479319e+00 - 5*1.25)-- (2.4549805e+01 - 4, 9.9479319e+00 - 5*1.25);
\draw[blue]  (2.4549805e+01 - 2.8, 9.9479319e+00 - 5*1.25)-- (2.4549805e+01 - 2.8, 9.9479319e+00 - 6*1.25);
\draw[blue]  (2.4549805e+01 - 0.4, 9.9479319e+00 - 6*1.25)-- (2.4549805e+01 - 2.8, 9.9479319e+00 - 6*1.25);
\draw[blue]  (2.4549805e+01 - 1.6, 9.9479319e+00 - 6*1.25)-- (2.4549805e+01 - 1.6, 9.9479319e+00 - 7*1.25);
\draw[blue]  (2.4549805e+01 - 0.4, 9.9479319e+00 - 7*1.25)-- (2.4549805e+01 - 1.6, 9.9479319e+00 - 7*1.25);

\filldraw[blue] (11,1) circle (2pt) node[anchor=south] {$r$};
\addplot [only marks, mark size = 2,black] table {plots/points/ptqam_points_polar_32_8_snr_25__zone_2_.txt};
\addplot [no markers, update limits=false] table{plots/voronoi/ptqam_voronoi_polar_32_8_snr_25__zone_2_.txt};
\end{axis}
\end{tikzpicture}	
\end{minipage}
\caption{Hexagonal decision regions formulated by GPD-D of a SAPSK($32$,$8$) constellation oriented a) along $\rho$ axis, b) along $\theta$ axis} 
\label{fig:PTQAM-decision-regions}
\end{figure}

\begin{remark}
As it can be seen in \eqref{proposed_detection_metric}, the GPD-D detection metric can be characterized a weighted Euclidean distance in the polar domain which produces symbol-independent and symmetrical decision regions based on the received symbol energy. Thus, this detector provides a foundation for the development of time-efficient nearest neighbor detection algorithms based on the constellation's geometry in the polar plane. This approach is aimed at addressing the demand for low-complexity detection in super-constellation communications. Consequently, in the context of SAPSK, by applying GPD-D hexagonal regions are formulated in the polar plane, thus SAPSK symbols are allocated in the polar plane in line with the optimal 2D sphere packing \cite{hqam_th}. 
\end{remark}

\section{Error Analysis and Detection in SAPSK}

As mentioned before, SAPSK stands out as a constellation with its symbols being compactly positioned in the polar plane, thereby providing a solution aligning with the criteria of super-constellation communications. Within this framework, the introduction of GPD-D, enables the analytical investigation of the SEP of SAPSK constellations and also lays the foundation for effective detection algorithms. 
In the following subsections, we characterize how the shape of the SAPSK decision regions is adjusted to the varying noise conditions, and based on that, we provide a tight closed-form approximation for their SEP and a novel detection algorithm with $\mathcal{O}\left(1\right)$ computational complexity.

\subsection{Performance Analysis of SAPSK Constellations}

To calculate the SEP of an SAPSK$\left(M, \Gamma\right)$ constellation under AWGN and GPN conditions, the shape of the decision regions that are formulated by GPD-D is required. As illustrated in Fig. \ref{fig:QAM-PTQAM-iq-polar}f, the $M$ symbols of an SAPSK$\left(M, \Gamma\right)$ constellation are arranged in $\Gamma$ columns in the polar plane, particularly allocated to form a triangular lattice. In this direction, as shown in Fig. \ref{fig:PTQAM-decision-regions}, by applying GPD-D, which coincides with applying nearest neighbour in the polar plane, the formulated decision regions of an SAPSK are equivalent to irregular hexagons. However, by taking into account \eqref{proposed_detection_metric}, the shape of the decision regions is affected by the values of AWGN power $\sigma_{n}^2$, the GPN level ${\sigma_{\phi}}^2$, and the energy of the received symbol $r$, signifying the alteration of the decision regions based on the prevailing noise conditions. In more detail, the polar coordinates of the constellation symbols and the received symbol $r$ undergo a scaling defined by the denominators in \eqref{proposed_detection_metric}, which results in the hexagonal decision regions being elongated along the $\rho$ and $\theta$ axes, and thus reorienting the hexagons along these axes, as illustrated in Fig. \ref{fig:PTQAM-decision-regions}. It should be highlighted, that the decision region reorientation is a mechanism of GPD-D to effectively mitigate the symbol shift induced by AWGN along $\rho$ and $\theta$ axes, as well as the shift induced by GPN along the $\theta$ axis depending on the energy of the received symbol $r$. 

By taking into account the shape of the decision regions, a SEP approximation for SAPSK constellations can be derived by decomposing the hexagonal regions into a rectangular region along with two triangles whose areas are further decomposed to a definite number of rectangles, as shown in Fig. \ref{fig:PTQAM-decision-regions}. However, based on the prevailing noise conditions the hexagons are correspondingly extended either along $\rho$ or $\theta$ axis, which leads to the variation of their shape. In more detail, in cases where the AWGN is greater than GPN, the hexagons are extended along $\rho$ axis to mitigate symbol displacements between two adjacent energy levels induced by AWGN, as illustrated in Fig. \ref{fig:PTQAM-decision-regions}a. Conversely, in cases where the GPN becomes the dominant distortion, the decision regions start to extend along $\theta$ axis, as shown in Fig. \ref{fig:PTQAM-decision-regions}b, to mitigate the displacements that occur only along the $\theta$ axis due to GPN.

Building upon the prevailing noise conditions, the shape of the hexagonal decision regions varies, and thus the location of the triangular regions that will be decomposed to rectangles changes as well. Therefore, to approximate the triangular regions with rectangles, we need to characterize the decision regions' orientation and calculate the received SNR threshold value $\gamma_q$ above which the hexagons are reoriented from $\rho$ to $\theta$ axis.
\begin{proposition}\label{prop:orientation} Assuming that $r$ lies between $q$-th and $\left(q+1\right)$-th energy levels and that the decision regions are designed based on GPD-D, the orientation $O_h$ of the hexagonal decision regions of an SAPSK$\left(M, \Gamma\right)$ constellation is given as
\begin{equation}\label{O_h}
\small
\begin{aligned}
    O_{h}\equiv 
    \begin{cases}
    \begin{split}
        \textbf{$\rho$ axis},
   \end{split}
      &\text{$\bar{\gamma}$} \leq \gamma_{q}\\\\
   \begin{split}
   \textbf{$\theta$ axis},
   \end{split} 
      &\text{$\bar{\gamma}$} > \gamma_{q},
   \end{cases}
\end{aligned}
\end{equation}
where $\gamma_q$ is a received SNR threshold value given as
\begin{equation}\label{gamma_q}
\begin{aligned}
  \gamma_{q} = \frac{\left(2q-1\right)^2\pi^2\Gamma^2\left(4\Gamma^2-1\right)-4M^2\left(4\Gamma^2-1\right)}{24M^2\left(2q-1\right)^2\sigma_{\phi}^2}.
\end{aligned}
\end{equation}
\end{proposition}
\begin{IEEEproof}
The proof is provided in Appendix \ref{proof_Prop2}.
\end{IEEEproof}
\begin{remark}
While GPD-D aims to adaptively extend the decision regions to enhance GPN resilience, the effectiveness of this adaptation is hindered under high GPN conditions. Specifically, as GPN severity increases, this strategic extension of the decision regions along the $\theta$ axis cannot reduce the error floor, since the elevated level of GPN leads to a high probability of the received symbol being displaced outside of its designated decision region, limiting the efficiency of GPD-D in compensating the impact of GPN. 
\end{remark}

Having calculated the threshold $\gamma_{q}$ under which the hexagonal decision regions reorient in the polar plane, the positions of the triangular sub-areas can be defined, and thus their decomposition to rectangles can be performed. To this end, in the following proposition, a tight closed-form approximation for the SEP of an SAPSK$\left(M, \Gamma\right)$ constellation is derived.
\begin{proposition}\label{prop:PTQAM}
By decomposing each hexagonal decision region with a main rectangle and two triangles that are approximated by $2N$ rectangles, the SEP of an SAPSK$\left(M, \Gamma\right)$ constellation affected by AWGN and GPN can be tightly approximated as
    \begin{equation} \label{SEP_ptqam_proposed_metric}
    \small
        \begin{aligned}
            P_e\approx\frac{1}{\Gamma}\sum_{q=1}^{\Gamma}P_q ,
            \end{aligned}
    \end{equation}
    where $P_q$ denotes the error probability of symbol $s_q$ that lies on the $q$-th energy level and it is written as in \eqref{Prob_q} at the top of the next page, and $Q(x) = \frac{1}{\sqrt{2\pi}} \int_x^\infty \exp\left(-\frac{u^2}{2}\right) \, du$ is the Gaussian $Q$-function. Furthermore, $B_z$ is given as 
\begin{figure*}

\begin{equation}\label{Prob_q}
\small
    \begin{aligned}
        P_{q} = 
        \begin{cases}
        \begin{split}
            &1-\left(1-2Q\left(\sqrt{\frac{24\bar{\gamma}}{2\Gamma^2-1}}\right)\right)\left(1-2Q\left(\frac{\left(1-D_q\right)\pi\Gamma}{M\sqrt{\pn^2 + \frac{\sigma_{n}^2}{2E_q}}}\right)\right) \\
&\quad -4\sum_{z=1}^{N}\left(Q\left(\frac{\left(1-D_q\right)\pi\Gamma}{M\sqrt{\pn^2 + \frac{\sigma_{n}^2}{2E_q}}}\right)-Q\left(\frac{B_z\pi\Gamma}{M\sqrt{\pn^2 + \frac{\sigma_{n}^2}{2E_q}}}\right)\right)\left(Q\left(\sqrt{\frac{24\bar{\gamma}(z-1)^2}{\left(4\Gamma^2-1\right)N^2}}\right)-Q\left(\sqrt{\frac{24\bar{\gamma}z^2}{\left(4\Gamma^2-1\right)N^2}}\right)\right),
       \end{split}
        &\text{$\bar{\gamma}$} \leq \gamma_q\\
       \begin{split}
       &1-\left(1-2Q\left(\sqrt{\frac{24\bar{\gamma}(1-D_q)^2}{4\Gamma^2-1}}\right)\right)\left(1-2Q\left(\frac{\pi\Gamma}{M\sqrt{\pn^2+\frac{\sigma_n^2}{2E_q}}}\right)\right)\\
&\quad -4\sum_{z=1}^{N}\left(Q\left(\sqrt{\frac{24\bar{\gamma}(1-D_q)^2}{4\Gamma^2-1}}\right)-Q\left(\sqrt{\frac{24\bar{\gamma}B_{z}}{4\Gamma^2-1}}\right)\right)\left(Q\left(\frac{(z-1)\pi\Gamma}{NM\sqrt{\pn^2+\frac{\sigma_n^2}{2E_q}}} \right)-Q\left( \frac{z\pi\Gamma}{NM\sqrt{\pn^2+\frac{\sigma_n^2}{2E_q}}}\right)\right),
       \end{split} 
       &\text{$\bar{\gamma}$} > \gamma_q
     \end{cases}
    \end{aligned}
\end{equation}
\hrule
\end{figure*}
\begin{equation}\label{Bz}
    \begin{aligned}
        B_z = \frac{\left(1-2D_q\right)z}{N} + D_q,
    \end{aligned}
\end{equation}
with $z\in\{1,\dots,N\}$, and $D_q$ is given by 
\begin{equation}\label{K}
\small
    \begin{aligned}
        D_q = 
        \begin{cases}
        \begin{split}
            \frac{24M^2\Bar{\gamma}\left(\sigma_{\phi}^2 + \frac{4\Gamma^2-1}{6\Bar{\gamma}(2q-1)^2}\right) + \pi^2\Gamma^2\left(4\Gamma^2-1\right)}{2\pi^2\Gamma^2\left(4\Gamma^2-1\right)},
       \end{split}
        &\text{$\bar{\gamma}$} \leq \gamma_q\\\\
       \begin{split}
       \frac{24M^2\bar{\gamma}\left(\sigma_{\phi}^2 + \frac{4\Gamma^2-1}{6\Bar{\gamma}(2q-1)^2}\right) + \pi^2\Gamma^2\left(4\Gamma^2-1\right)}{48M^2\bar{\gamma}\left(\sigma_{\phi}^2 + \frac{4\Gamma^2-1}{6\Bar{\gamma}(2q-1)^2}\right)},
       \end{split} 
       &\text{$\bar{\gamma}$} > \gamma_q.
     \end{cases}
    \end{aligned}
\end{equation}
\end{proposition}

\begin{IEEEproof}
The proof is provided in Appendix \ref{proof_Prop3}.
\end{IEEEproof}
\begin{remark}
Considering a fixed $M$, SNR, and $\pn^2$, \eqref{SEP_ptqam_proposed_metric} provides a straightforward method for obtaining the optimal design parameter $\Gamma$ of an SAPSK$\left(M, \Gamma\right)$, which minimizes the SEP for a given energy constraint, thereby enhancing the overall energy efficiency of the system.
\end{remark}
\begin{remark}
As $\bar{\gamma}\rightarrow\infty$ the $P_{e}$ is equal to
\begin{equation}
    \begin{aligned}
        P_e = Q\left(\frac{2\pi\Gamma}{M\pn}\right).
    \end{aligned}
\end{equation}
\end{remark}

\subsection{SAPSK Detection Algorithm for practical implementation in super-constellation communications}
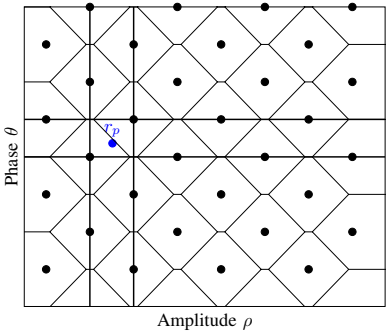
\begin{figure}[ht!]
\centering
\begin{tikzpicture}[scale=0.7]
\begin{axis}[
    xtick=\empty,
    ytick=\empty,
    xlabel={$\text{Amplitude $\rho$}$},
    ylabel={$\text{Phase $\theta$}$},
    xlabel style={below},
    ylabel style={above},
    xmin=0,
    xmax=45,
    ymin=-19.85,
    ymax=19.9,
]
\draw[black,thick] (8.183,-19.85) -- (8.183,19.9);
\draw[black,thick] (13.639,-19.85) -- (13.639,19.9);
\draw[black,thick] (0,0) -- (45,0);
\draw[black,thick] (0,4.973966) -- (45,4.973966);
\filldraw[blue] (11,1.8) circle (2pt) node[anchor=south] {$r_{p}$};
\addplot [only marks, mark size = 2,black] table {plots/points/ptqam_points_polar_32_8_snr_25__zone_2_.txt};
\addplot [no markers, update limits=false] table{plots/voronoi/ptqam_voronoi_polar_32_8_snr_25__zone_2_.txt};
\end{axis}
\end{tikzpicture}	
\caption{Detection scheme}
\label{fig:detection_scheme}
\end{figure}

In the context of an SAPSK constellation affected by AWGN and GPN, the decision rule outlined in (\ref{optimal_detection_metric}) stands as the optimal maximum likelihood detection method, effectively minimizing symbol error probability. Despite its theoretical optimality, this approach describes a linear traversal of all the symbols in the constellation to detect the received signal, resulting in a complexity of $\mathcal{O}(M)$. Such an algorithmic complexity deteriorates the system performance in terms of time, especially in super constellations where $M$ takes large values. Conversely, the GPD-D metric offers a promising avenue for developing time-efficient algorithms with lower complexity, while still preserving demodulation accuracy compared to the optimal detector. Specifically, this decision rule can be conceptualized as a joint amplitude and phase detector, entailing a weighted combination of amplitude and phase differences between the received signal and the symbols of the constellation within specific zones in the polar domain. To this end, in this section a novel detection algorithm for SAPSK constellations of complexity $\mathcal{O}(1)$ is presented.

Let the $i$-th symbol of constellation $\mathcal{C}$ be of the form $s_{i} = x_{i}+jy_{i}$, where $j$ is the imaginary unit. Each symbol is written in the polar form as $s_{i} = \vert s_{i}\vert e^{j\arg\{s_{i}\}}$ and it is one-to-one mapped to a symbol in the polar domain of the form $s_{\mathrm{p}i} = \vert s_{i}\vert +j\arg\{s_{i}\}$, where $\vert s_{i}\vert\in[0,\infty)$ and $\arg\{s_{i}\}\in[-\pi,\pi]$. Furthermore, after receiving a symbol $r = x_r + jy_r = \lvert r\rvert e^{j\arg{r}}$, it is transformed into the polar domain as $r_p = \vert r_{p}\vert + j\arg\{r_{p}\}$, as depicted in Fig. \ref{fig:detection_scheme}.
The key idea of our algorithm lies in identifying potential constellation symbols close to the received symbol $r_{p}$, thereby reducing the necessary number of computed Euclidean distances between $r$ and the constellation symbols. Within this framework, we define the regions $R_{1} = [\sqrt{E_{q}}, \sqrt{E_{q+1}}]$ and $R_{2} = [\theta_{w}, \theta_{w+1}]$, $q\in\{1,\dots,\Gamma\}$ and $w\in\{1,\dots,\frac{M}{\Gamma}\}$. Furthermore, $E_{q}$, $E_{q+1}$, $\theta_{w}$, and $\theta_{w+1}$ are the lines that are parallel to the axes, as illustrated in Fig. \ref{fig:detection_scheme}, and they enclose the received symbol $r_{p}$ between two consecutive energy levels and two consecutive angles of SAPSK constellation, respectively. The detection process is completed by obtaining the symbol $\hat{s} \in G = R_{1} \cap R_{2}$ which minimizes the metric, given by \eqref{proposed_detection_metric}. To accomplish this, it becomes evident that we need to identify the constellation symbols that are contained in the region $G$. In this direction, the $\Gamma$ distinct energy levels of the SAPSK constellation are stored in ascending order in the array $S_{E}$ as 
\begin{equation}\label{S_rho}
    S_{E} = [E_{1}, \dots,E_q,E_{q+1},\dots, E_{\Gamma}].
\end{equation}
Afterwards, for each $E_{q}\in S_{E}$ an adjacent list with all distinct $\theta$ values in this amplitude is created as 
\begin{equation}\label{A_Eq}
    \begin{aligned}
        A_{E_{q}} = [\theta_{q,1}, \theta_{q,2},\dots,\theta_{q,\frac{M}{\Gamma}}],
    \end{aligned}
\end{equation}
where $\theta$ values are sorted in ascending order.

To design an $\mathcal{O}(1)$ detection algorithm for SAPSK constellations, it is essential to exploit their structure and, particularly, the arrays $S_{E}$ and $A_{E_q}$. In this direction, by noting that for an SAPSK constellation the values in $S_{E}$ and $A_{E_q}$ increase with a constant rate, we can formulate a linear interpolation function $f(\cdot)$ that returns the position of its input in the arrays, and thus find the constellation symbols within $G$ in $\mathcal{O}(1)$ time complexity. In particular, when the input value $u$ of $f$ matches one of the array elements, $f$ returns an integer value representing the position of $u$ in the array. Conversely, if $u$ is not an array element, $f$ yields a non-integer value, which results in the position of $S_{E}$ with the closest corresponding value to $u$ when rounded. Therefore, given that the elements of $S_{E}$ are equispaced by $\delta_{\rho}$ due to the SAPSK geometry, we can define the linear interpolation function $f_{E}: \mathbb{R} \rightarrow \mathbb{R}$, which is given by
\begin{equation}\label{inter_x}
    \begin{aligned}
        f_{E}(u) = \frac{1}{\delta_{\rho}}v + 1-\frac{\sqrt{E_{1}}}{\delta_{\rho}},
    \end{aligned}
\end{equation}
where $E_1$ is the first element of $S_{E}$. To this end, by utilizing \eqref{inter_x}, we can obtain the consecutive positions $q$ and $q+1$ in $S_{E}$ that satisfy $\sqrt{E_q} \leq \lvert r\rvert \leq \sqrt{E_{q+1}}$, which are given as
\begin{equation} \label{eq:xi}
    \begin{aligned}
         &q = \floor*{f_{E}(|r|)} \\
         &q+1 = \ceil*{f_{E}(|r|)},
    \end{aligned}
\end{equation}
where $\ceil*{\cdot}$ and $\floor*{\cdot}$ are the ceil function. Similarly, given that the elements of $A_{E_q}$ are equispaced by $\delta_{\theta}$ due to the SAPSK geometry, we can also define the linear interpolation function $f_{\theta,q}: \mathbb{R} \rightarrow \mathbb{R}$, which is given by
\begin{equation}\label{inter_Ax}
    \begin{aligned}
        f_{\theta,q}(v) = \frac{1}{\delta_{\theta}}v + 1-\frac{\theta_{q,1}}{\delta_{\theta}},
    \end{aligned}
\end{equation}
where $\theta_{q,1}$ is the first element of $A_{E_q}$. Therefore, the consecutive positions $\left(w, w+1\right)$ and $\left(q, q+1\right)$ of $A_{E_q}$ and $A_{E_{q+1}}$, respectively, that satisfy $\theta_{q, w} \leq \arg\{r\} \leq \theta_{q, w+1}$ and $\theta_{q+1,l} \leq \arg\{r\} \leq \theta_{q+1, l+1}$ can be obtained as \color{black}
\begin{equation} \label{eq:j}
    \begin{aligned}
        &w = \floor*{f_{\theta,q}(\arg\{r\})}\\
        &w+1 = \ceil*{f_{\theta,q}(\arg\{r\})}\\
        &l = \floor*{f_{\theta,q+1}(\arg\{r\})}\\
        &l+1 = \ceil*{f_{\theta,q+1}(\arg\{r\})}.
    \end{aligned}
\end{equation}
Consequently, we can obtain the set $S_c$ that contains the symbols of SAPSK on region $G$ which can be expressed in IQ coordinates as 
\begin{equation}
\small
\begin{aligned}
    S_c = 
    \begin{cases}
    \begin{split}
        \left\{\sqrt{E_q} e^{j\theta_{q,w}}, \sqrt{E_{q+1}} e^{j\theta_{q+1,l+1}}\right\},
   \end{split}
    &\text{if $q$ even}\\\\
   \begin{split}
    \left\{\sqrt{E_q} e^{j\theta_{q,w+1}}, \sqrt{E_{q+1}} e^{j\theta_{q+1,l}}\right\},
   \end{split} 
   &\text{if $q$ odd}.
 \end{cases}
\end{aligned}
\end{equation}

Finally, the symbol $\hat{s}$ that GPD-D determines as the most likelihood to have been sent can be obtained as
\begin{equation}
    \begin{aligned}
         \hat{s} = \arg \min_{s\in S_c}\hspace{4px}\frac{\big(\lvert r \rvert-\lvert s \rvert \big)^2}{\sigma_{n}^2} + \frac{\big(\arg\{r\}-\arg\{s\}\big)^2}{\sigma_{\phi}^2 + \frac{\sigma_{n}^2}{2\lvert r\rvert^2}},
    \end{aligned}
\end{equation}
It should be highlighted that for any $M$-ary SAPSK constellation, it holds that $\vert S_c \vert \leq 2$. Consequently, instead of comparing $M$ symbols as required by the optimal detector in the presence of GPN, known as GAP-D, we only need to perform two calculations to determine which symbol was transmitted, resulting in $\mathcal{O}(1)$ complexity.

\begin{algorithm}
    \caption{SAPSK Constellation Detection Algorithm}
    \begin{algorithmic}[1]
        \raggedright
        \renewcommand{\algorithmicrequire}{\textbf{Input:}}
        \renewcommand{\algorithmicensure}{\textbf{Output:}}
        \REQUIRE Coordinates of the received symbol $r = x_r + jy_r$
        \ENSURE  Detected symbol $\hat{s}$
        \STATE Transform $r$ into polar coordinates: $r_p = \lvert r \rvert + j\arg\{r\}$
        \STATE Initialize arrays $S_{E}$ for storing energy levels
        \STATE Initialize arrays $A_{E_q}$, $\forall E_q\in S_{E}$, for storing the symbol arguments on each energy level
        \STATE Calculate $q$-th and $\left(q+1\right)$-th energy levels using linear interpolation: $q = \floor*{f_{E}(\lvert r \rvert)}$, $q+1 = \ceil*{f_{E}(\lvert r \rvert)}$
        \STATE Calculate $w$-th and $\left(w+1\right)$-th symbol arguments of the $q$-th energy level using linear interpolation within $A_{E_q}$: $w = \floor*{f_{\theta,q}(\arg\{r\})}$, $w+1 = \ceil*{f_{\theta,q}(\arg\{r\})}$
        \STATE Calculate $l$-th and $\left(l+1\right)$-th symbol arguments of the $\left(q+1\right)$-th energy level  using linear interpolation within $A_{E_{q+1}}$: $l = \floor*{f_{\theta,q+1}(\arg\{r\})}$, $l+1 = \ceil*{f_{\theta,q+1}(\arg\{r\})}$
        \STATE Construct the candidate set
        \begin{equation*}
        \small
        \begin{aligned}
            S_c = 
            \begin{cases}
            \begin{split}
                \left\{\sqrt{E_q} e^{j\theta_{q,w}}, \sqrt{E_{q+1}} e^{j\theta_{q+1,l+1}}\right\},
           \end{split}
            &\textbf{$q$ even}\\\\
           \begin{split}
            \left\{\sqrt{E_q} e^{j\theta_{q,w+1}}, \sqrt{E_{q+1}} e^{j\theta_{q+1,l}}\right\},
           \end{split} 
           &\textbf{$q$ odd}.
         \end{cases}
        \end{aligned}
        \end{equation*}
        \STATE Find the symbol $\hat{s}$ in $S_c$ that minimizes the metric: $\hat{s} = \arg \min_{s\in S_c}\frac{(\lvert r \rvert-\lvert s \rvert)^2}{\sigma_{n}^2} + \frac{(\arg\{r\}-\arg\{s\})^2}{\sigma_{\phi}^2 + \frac{\sigma_{n}^2}{2\lvert r\rvert^2}}$
        \RETURN $\hat{s}$
    \end{algorithmic}
    \label{det_algo}
\end{algorithm}
\section{Numerical Results}
In this section, we verify the effectiveness of the proposed GPD-D in channels affected by both AWGN and GPN, as well as the accuracy of the proposed SEP approximation for an SAPSK$\left(M, \Gamma\right)$ constellation, where the trade-off involving SEP, SNR, $\pn^2$, $M$, and $\Gamma$ is illustrated. Furthermore, we show the gain in terms of energy efficiency that we obtain when SAPSK is used over PQAM in super-constellation communications. Finally, we validate the detection accuracy of the proposed detection algorithm for SAPSK constellations whose time complexity is independent of $M$ and achieves a detection performance equivalent to GAP-D. 

\begin{figure}[h!]
\centering
\begin{minipage}{.5\textwidth}
\centering
\begin{tikzpicture}
    \begin{semilogyaxis}[
        width=0.78\linewidth,
	xlabel = $\frac{E_{s}}{N_{0}}$ (dB),
	ylabel = Symbol Error Probability,
	xmin = 30,
	xmax = 80,
	ymin = 0.000001,
	ymax = 1,
        xtick = {0,10,...,80},
	grid = major,
        legend cell align = {left},
        legend pos = south west,
        legend style={font=\tiny}
    ]
    \addplot[
        color= black,
        only marks,
	mark=*,
	mark repeat = 2,
	mark size = 2,
    ]
	table {plots/transaction-data/pqam/4096-PQAM_512_-sim-proposed-metric-_1.000000e-04_.txt};
    \addlegendentry{4096-PQAM(512) \eqref{proposed_detection_metric}}
    
    \addplot[
        color= blue,
        only marks,
	mark=diamond*,
	mark repeat = 2,
	mark size = 3,
    ]
	table {plots/transaction-data/ptqam/4096-PTQAM_512_-sim-proposed-metric-_1.000000e-04_.txt};
    \addlegendentry{SAPSK(4096,512) \eqref{proposed_detection_metric}}
    
    \addplot[
        color= brown,
        only marks,
	mark=square*,
	mark repeat = 2,
	mark size = 2,
    ]
	table {plots/transaction-data/qam/4096-QAM-sim-propd_1.000000e-04_smoothed.txt};
    \addlegendentry{4096-QAM \eqref{proposed_detection_metric}}

    \addplot[
        no marks,
	mark repeat = 2,
	mark size = 1.5,
        line width = 1pt,
        style = solid
    ]
	table {plots/transaction-data/pqam/4096-PQAM_512_-sim-OPT-metric-_1.000000e-04_.txt};
    \addlegendentry{GAPD}
    
    \addplot[
        color = blue,
        no marks,
	mark repeat = 2,
	mark size = 2,
        line width = 1pt,
        style = solid
    ]
	table {plots/transaction-data/ptqam/4096-PTQAM_512_-sim-OPT-metric-_1.000000e-04_.txt};

    \addplot[
        color = brown,
	no marks,
	mark repeat = 2,
	mark size = 2,
        line width = 1pt,
        style=solid
    ]
	table {plots/transaction-data/qam/4096-QAM-sim-propd_1.000000e-04_smoothed.txt};

    \end{semilogyaxis}
\end{tikzpicture}
\subcaption{$\sigma^2_{\phi}= 10^{-4}$}
\end{minipage}
\begin{minipage}{.5\textwidth}
\centering
\begin{tikzpicture}
    \begin{semilogyaxis}[
        width=0.78\linewidth,
	xlabel = $\frac{E_{s}}{N_{0}}$ (dB),
	ylabel = Symbol Error Probability,
	xmin = 30,
	xmax = 80,
	ymin = 0.00001,
	ymax = 1,
        xtick = {0,10,...,80},
	grid = major,
        legend cell align = {left},
        legend pos = south west,
        legend style={font=\tiny}
    ]

    \addplot[
        color= black,
        only marks,
	mark=*,
	mark repeat = 2,
	mark size = 2,
    ]
	table {plots/transaction-data/pqam/4096-PQAM_512_-sim-proposed-metric-_1.000000e-02_.txt};
    \addlegendentry{4096-PQAM(512) \eqref{proposed_detection_metric}}
    \addplot[
        color= blue,
        only marks,
	mark=diamond*,
	mark repeat = 2,
	mark size = 3,
    ]
	table {plots/transaction-data/ptqam/4096-PTQAM_512_-sim-proposed-metric-_1.000000e-02_.txt};
    \addlegendentry{SAPSK(4096,512) \eqref{proposed_detection_metric}}
    \addplot[
        color= brown,
        only marks,
	mark=square*,
	mark repeat = 2,
	mark size = 2,
    ]
	table {plots/transaction-data/qam/4096-QAM-sim-GAPD_1.000000e-02_.txt};
    \addlegendentry{4096-QAM \eqref{proposed_detection_metric}}

    \addplot[
        no marks,
	mark repeat = 2,
	mark size = 1.5,
        line width = 1pt,
        style = solid
    ]
	table {plots/transaction-data/pqam/4096-PQAM_512_-sim-OPT-metric-_1.000000e-02_.txt};
    \addlegendentry{GAPD}
    
    \addplot[
        color = blue,
        no marks,
	mark repeat = 2,
	mark size = 2,
        line width = 1pt,
        style = solid
    ]
	table {plots/transaction-data/ptqam/4096-PTQAM_512_-sim-OPT-metric-_1.000000e-02_.txt};

    \addplot[
        color = brown,
	no marks,
	mark repeat = 2,
	mark size = 2,
        line width = 1pt,
        style=solid
    ]
	table {plots/transaction-data/qam/4096-QAM-sim-propd_1.000000e-02_.txt};

    \end{semilogyaxis}
\end{tikzpicture}
\subcaption{$\sigma^2_{\phi}= 10^{-2}$}
\end{minipage}
\caption{Performance of GPD-D metric for various constellation design}
\label{fig:metrics_comparison}
\end{figure}
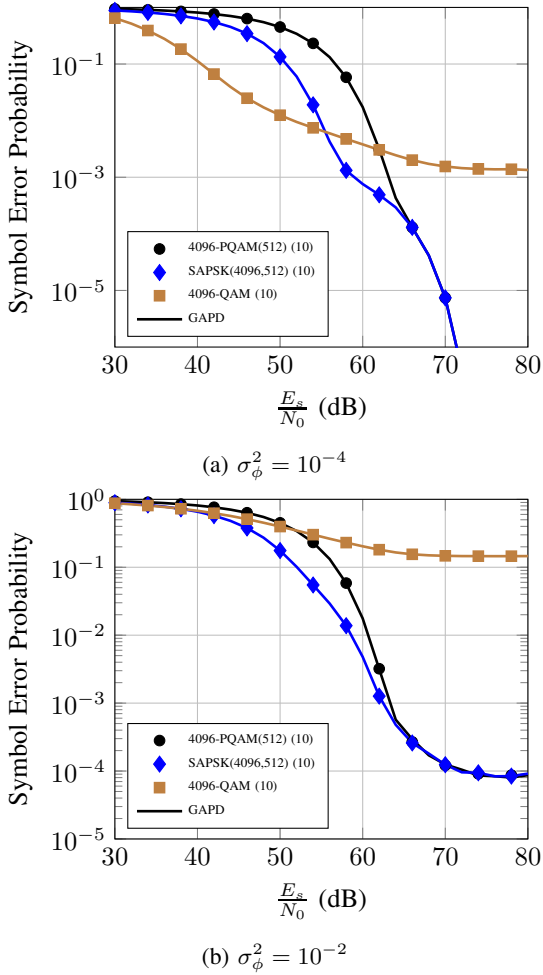

In Figs. \ref{fig:metrics_comparison}a and \ref{fig:metrics_comparison}b, it is illustrated that GPD-D achieves equivalent performance to the optimal GAP-D metric under two GPN scenarios for different constellation designs. In particular, it can be seen that in both considered GPN scenarios, i.e., $\pn^2=10^{-4}$ and $\pn^2=10^{-2}$, the performance of GPD-D is identical to GAP-D, indicating its effectiveness for different GPN conditions. In addition, similar to GAP-D as shown in Fig. \ref{fig:euc_vs_gap}, the system performance is severely affected by a high error floor when GPD-D and QAM are used, highlighting a significant degradation in super-constellation communications when constellations not tailored to the existence of GPN are used. However, when modulation schemes such as PQAM and SAPSK are employed along with GPD-D, the system exhibits significantly improved performance, highlighting the importance of novel GPN-resilient modulation schemes for super-constellation communications, even at relatively low GPN levels, to reduce the incurred error floor and thus ensure reliable SEP performance.

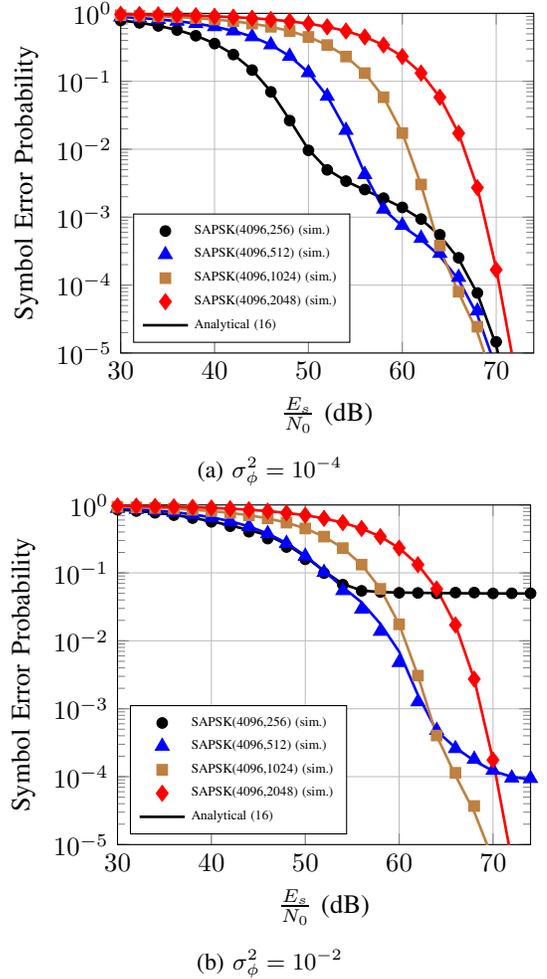
\begin{figure}[h!]
\centering
\begin{minipage}{.5\textwidth}
\centering
\begin{tikzpicture}
	\begin{semilogyaxis}[
    width=0.78\linewidth,
	xlabel = $\frac{E_{s}}{N_{0}}$ (dB),
	ylabel = Symbol Error Probability,
	xmin = 30,
	xmax = 74,
	ymin = 0.00001,
	ymax = 1,
    xtick = {0,10,...,80},
	grid = major,
    legend cell align = {left},
    legend pos = south west,
    legend style={font=\tiny}
	]

    \addplot[
        color= black,
        only marks,
	mark=*,
	mark repeat = 1,
	mark size = 2,
    ]
	table {plots/transaction-data/ptqam/4096-PTQAM_256_-sim-OPT-metric-_1.000000e-04_.txt};
    \addlegendentry{SAPSK(4096,256) (sim.)}
    
    \addplot[
        color= blue,
        only marks,
	mark=triangle*,
	mark repeat = 1,
	mark size = 3,
    ]
	table {plots/transaction-data/ptqam/4096-PTQAM_512_-sim-OPT-metric-_1.000000e-04_.txt};
    \addlegendentry{SAPSK(4096,512) (sim.)}
 
    \addplot[
        color = brown,
        only marks,
	mark=square*,
	mark repeat = 1,
	mark size = 2,
    ]
	table {plots/transaction-data/ptqam/4096-PTQAM_1024_-sim-OPT-metric-_1.000000e-04_.txt};
    \addlegendentry{SAPSK(4096,1024) (sim.)}
 
    \addplot[
        color= red,
        only marks,
	mark=diamond*,
	mark repeat = 1,
	mark size = 3,
    ]
	table {plots/transaction-data/ptqam/4096-PTQAM_2048_-sim-OPT-metric-_1.000000e-04_.txt};
    \addlegendentry{SAPSK(4096,2048) (sim.)}
    
    \addplot[
	color = black,
        no marks,
	line width = 1pt,
	style = solid,
    ]
	table {plots/transaction-data/ptqam/4096-PTQAM_256_-theory-proposed-metric-_1.000000e-04_.txt};
    \addlegendentry{Analytical \eqref{SEP_ptqam_proposed_metric}}
    
    \addplot[
	color = blue,
        no marks,
	line width = 1pt,
	style = solid,
    ]
	table {plots/transaction-data/ptqam/4096-PTQAM_512_-theory-proposed-metric-_1.000000e-04_.txt};

    \addplot[
	color = brown,
        no marks,
	line width = 1pt,
	style = solid,
    ]
	table {plots/transaction-data/ptqam/4096-PTQAM_1024_-theory-proposed-metric-_1.000000e-04_.txt};

    \addplot[
	color = red,
        no marks,
	line width = 1pt,
	style = solid,
    ]
	table {plots/transaction-data/ptqam/4096-PTQAM_2048_-theory-proposed-metric-_1.000000e-04_.txt};
    \end{semilogyaxis}
\end{tikzpicture}
\subcaption{$\sigma^2_{\phi} = 10^{-4}$}
\end{minipage}
\begin{minipage}{.5\textwidth}
\centering
\begin{tikzpicture}
\begin{semilogyaxis}[
width=0.78\linewidth,
xlabel = $\frac{E_{s}}{N_{0}}$ (dB),
ylabel = Symbol Error Probability,
xmin = 30,
xmax = 74,
ymin = 0.00001,
ymax = 1,
xtick = {0,10,...,70},
grid = major,
legend cell align = {left},
legend pos = south west,
legend style={font=\tiny}
]
    \addplot[
        color= black,
        only marks,
	mark=*,
	mark repeat = 1,
	mark size = 2,
    ]
	table {plots/transaction-data/ptqam/4096-PTQAM_256_-sim-OPT-metric-_1.000000e-02_.txt};
    \addlegendentry{SAPSK(4096,256) (sim.)}
    
    \addplot[
        color= blue,
        only marks,
	mark=triangle*,
	mark repeat = 1,
	mark size = 3,
    ]
	table {plots/transaction-data/ptqam/4096-PTQAM_512_-sim-OPT-metric-_1.000000e-02_.txt};
    \addlegendentry{SAPSK(4096,512) (sim.)}
 
    \addplot[
        color = brown,
        only marks,
	mark=square*,
	mark repeat = 1,
	mark size = 2,
    ]
	table {plots/transaction-data/ptqam/4096-PTQAM_1024_-sim-OPT-metric-_1.000000e-02_.txt};
    \addlegendentry{SAPSK(4096,1024) (sim.)}
 
    \addplot[
        color= red,
        only marks,
	mark=diamond*,
	mark repeat = 1,
	mark size = 3,
    ]
	table {plots/transaction-data/ptqam/4096-PTQAM_2048_-sim-OPT-metric-_1.000000e-02_.txt};
    \addlegendentry{SAPSK(4096,2048) (sim.)}
    
    \addplot[
	color = black,
        no marks,
	line width = 1pt,
	style = solid,
    ]
	table {plots/transaction-data/ptqam/4096-PTQAM_256_-theory-proposed-metric-_1.000000e-02_.txt};
    \addlegendentry{Analytical \eqref{SEP_ptqam_proposed_metric}}
    
    \addplot[
	color = blue,
        no marks,
	line width = 1pt,
	style = solid,
    ]
	table {plots/transaction-data/ptqam/4096-PTQAM_512_-theory-proposed-metric-_1.000000e-02_.txt};

    \addplot[
	color = brown,
        no marks,
	line width = 1pt,
	style = solid,
    ]
	table {plots/transaction-data/ptqam/4096-PTQAM_1024_-theory-proposed-metric-_1.000000e-02_.txt};

    \addplot[
	color = red,
        no marks,
	line width = 1pt,
	style = solid,
    ]
	table {plots/transaction-data/ptqam/4096-PTQAM_2048_-theory-proposed-metric-_1.000000e-02_.txt};
\end{semilogyaxis}
\end{tikzpicture}
\subcaption{$\sigma^2_{\phi} = 10^{-2}$}
\end{minipage}
\caption{SAPSK SEP approximation}
\label{fig:approx}
\end{figure}

Figs. \ref{fig:approx}a and \ref{fig:approx}b validate the proposed SEP approximation for the SAPSK constellation for the case where $\pn^2=10^{-4}$ and $\pn^2=10^{-2}$, respectively. Specifically, the proposed SEP approximation is compared with the simulated SEP for the case of a 4096-SAPSK in the presence of GPN, for various $\Gamma$ values. As it can be seen, the tightness of the proposed approximation is confirmed for all investigated $\Gamma$ values, indicating the effectiveness of our approximation. Furthermore, in the low-SNR regime, where AWGN is the dominant distortion, smaller values of $\Gamma$ exhibit better SEP performance due to their property to mitigate the distortions caused by AWGN. However, as the SNR increases and the GPN becomes the dominant distortion, small values of $\Gamma$ show a performance degradation compared to higher values of $\Gamma$, because higher $\Gamma$ translates into assigning fewer symbols per energy level, thus increasing the robustness to the GPN. To this end, it can be deduced that there exists an optimal $\Gamma$ value according to the prevailing noise conditions and the order of the SAPSK constellation, highlighting the existence of a unique trade-off between energy efficiency and reliability in super-constellation communications. 

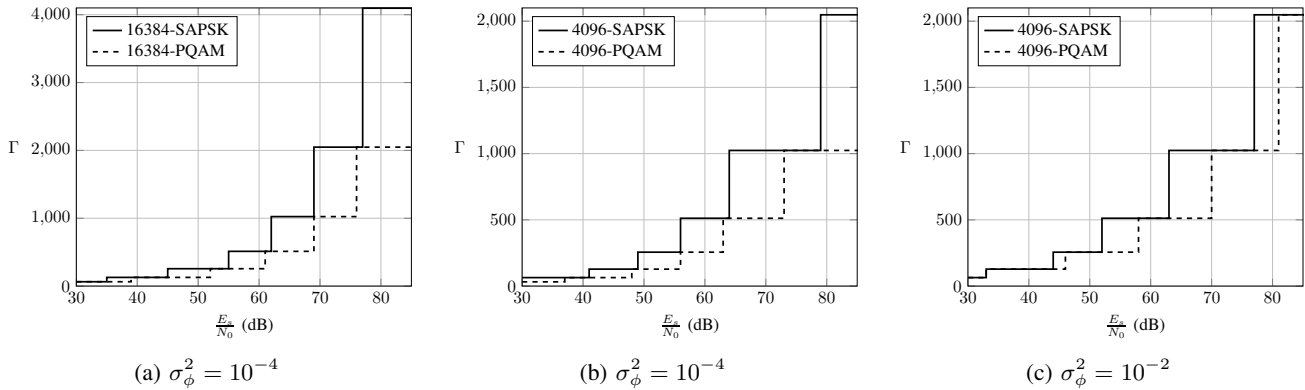
\begin{figure*}[t]
\centering
\begin{minipage}[t]{.32\textwidth}
\centering
\begin{tikzpicture}[scale = 0.65]
\begin{axis}[
xlabel = $\frac{E_{s}}{N_{0}}$ (dB),
ylabel = $\Gamma$,
xmin = 30,
xmax = 85,
ymin = 0,
ymax = 4100,
xtick = {0,10,...,100},
grid = major,
legend cell align = {left},
legend pos = north west,
ylabel style={rotate = -90}
]
\addplot[
no marks,
line width = 1pt,
style = solid,
]
table {plots/transaction-data/optG/16384-PTQAM_1.000000e-04_.txt};
\addlegendentry{16384-SAPSK}

\addplot[
no marks,
line width = 1pt,
style = dashed,
]
table {plots/transaction-data/optG/16384-PQAM_1.000000e-04_.txt};
\addlegendentry{16384-PQAM}
\end{axis}
\end{tikzpicture}
\subcaption{$\sigma^2_{\phi} = 10^{-4}$}
\end{minipage}
\begin{minipage}[t]{.32\textwidth}
\centering
\begin{tikzpicture}[scale = 0.65]
\begin{axis}[
xlabel = $\frac{E_{s}}{N_{0}}$ (dB),
ylabel = $\Gamma$,
xmin = 30,
xmax = 85,
ymin = 0,
ymax = 2100,
xtick = {0,10,...,80},
grid = major,
legend cell align = {left},
legend pos = north west,
ylabel style={rotate = -90}
]
\addplot[
no marks,
line width = 1pt,
style = solid,
]
table {plots/transaction-data/optG/4096-PTQAM_1.000000e-04_.txt};
\addlegendentry{4096-SAPSK}

\addplot[
no marks,
line width = 1pt,
style = dashed,
]
table {plots/transaction-data/optG/4096-PQAM_1.000000e-04_.txt};
\addlegendentry{4096-PQAM}
\end{axis}
\end{tikzpicture}
\subcaption{$\sigma^2_{\phi} = 10^{-4}$}
\end{minipage}
\begin{minipage}[t]{.32\textwidth}
\centering
\begin{tikzpicture}[scale = 0.65]
\begin{axis}[
xlabel = $\frac{E_{s}}{N_{0}}$ (dB),
ylabel = $\Gamma$,
xmin = 30,
xmax = 85,
ymin = 0,
ymax = 2100,
xtick = {0,10,...,80},
grid = major,
legend cell align = {left},
legend pos = north west,
ylabel style={rotate = -90}
]
\addplot[
no marks,
line width = 1pt,
style = solid,
]
table {plots/transaction-data/optG/4096-PTQAM_1.000000e-02_.txt};
\addlegendentry{4096-SAPSK}

\addplot[
no marks,
line width = 1pt,
style = dashed,
]
table {plots/transaction-data/optG/4096-PQAM_1.000000e-02_.txt};
\addlegendentry{4096-PQAM}
\end{axis}
\end{tikzpicture}
\subcaption{$\sigma^2_{\phi} = 10^{-2}$}
\end{minipage}
\caption{Optimal $\Gamma$ vs SNR}
\label{fig:optimal G}
\end{figure*}

\begin{figure}[h!]
\centering
\begin{minipage}{.5\textwidth}
\centering
\begin{tikzpicture}
	\begin{semilogyaxis}[
    width=0.78\linewidth,
	xlabel = $\frac{E_{s}}{N_{0}}$ (dB),
	ylabel = Symbol Error Probability,
	xmin = 30,
	xmax = 85,
	ymin = 0.00001,
	ymax = 1,
    xtick = {0,10,...,80},
	grid = major,
    legend cell align = {left},
    legend pos = south west,
    legend style={font=\tiny}
	]
\addplot[
color = black,
only marks,
mark = square*,
mark repeat = 5,
mark size = 2
]
table {plots/transaction-data/optG/SEP-4096-PTQAM_1.000000e-04_.txt};
\addlegendentry{4096-SAPSK (sim.)}

\addplot[
color = blue,
only marks,
mark = *,
mark repeat = 5,
mark size = 2
]
table {plots/transaction-data/optG/SEP-16384-PTQAM_1.000000e-04_.txt};
\addlegendentry{16384-SAPSK (sim.)}

\addplot[
color = black,
no marks,
line width = 1pt,
style = solid,
]
table {plots/transaction-data/optG/SEP-4096-PTQAM_1.000000e-04_.txt};
\addlegendentry{$M$-SAPSK (Analytical)}

\addplot[
color = black,
no marks,
line width = 1pt,
style = dashed,
]
table {plots/transaction-data/optG/SEP-4096-PQAM_1.000000e-04_.txt};
\addlegendentry{$M$-PQAM}
\addplot[
color = blue,
no marks,
line width = 1pt,
style = solid,
]
table {plots/transaction-data/optG/SEP-16384-PTQAM_1.000000e-04_.txt};

\addplot[
color = blue,
no marks,
line width = 1pt,
style = dashed,
]
table {plots/transaction-data/optG/SEP-16384-PQAM_1.000000e-04_.txt};
\end{semilogyaxis}
\end{tikzpicture}
\subcaption{$\sigma^2_{\phi} = 10^{-4}$}
\end{minipage}
\begin{minipage}{.5\textwidth}
\centering
\begin{tikzpicture}
\begin{semilogyaxis}[
width=0.78\linewidth,
xlabel = $\frac{E_{s}}{N_{0}}$ (dB),
ylabel = Symbol Error Probability,
xmin = 30,
xmax = 85,
ymin = 0.00001,
ymax = 1,
xtick = {0,10,...,85},
grid = major,
legend cell align = {left},
legend pos = south west,
legend style={font=\tiny}
]
\addplot[
color = black,
only marks,
mark = square*,
mark repeat = 5,
mark size = 2
]
table {plots/transaction-data/optG/SEP-4096-PTQAM_1.000000e-02_.txt};
\addlegendentry{4096-SAPSK (sim.)}

\addplot[
color = blue,
only marks,
mark = *,
mark repeat = 5,
mark size = 2
]
table {plots/transaction-data/optG/SEP-16384-PTQAM_1.000000e-02_.txt};
\addlegendentry{16384-SAPSK (sim.)}

\addplot[
color = black,
no marks,
line width = 1pt,
style = solid,
]
table {plots/transaction-data/optG/SEP-4096-PTQAM_1.000000e-02_.txt};
\addlegendentry{$M$-SAPSK (Analytical)}

\addplot[
color = black,
no marks,
line width = 1pt,
style = dashed,
]
table {plots/transaction-data/optG/SEP-4096-PQAM_1.000000e-02_.txt};
\addlegendentry{$M$-PQAM}
\addplot[
color = blue,
no marks,
line width = 1pt,
style = solid,
]
table {plots/transaction-data/optG/SEP-16384-PTQAM_1.000000e-02_.txt};

\addplot[
color = blue,
no marks,
line width = 1pt,
style = dashed,
]
table {plots/transaction-data/optG/SEP-16384-PQAM_1.000000e-02_.txt};
\end{semilogyaxis}
\end{tikzpicture}
\subcaption{$\sigma^2_{\phi} = 10^{-2}$}
\end{minipage}
\caption{SAPSK vs PQAM for optimal $\Gamma$}
\label{fig:PTQAvsPQAM_opt_G}
\end{figure}
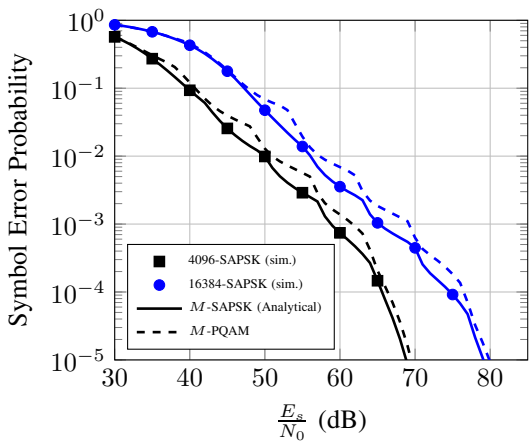
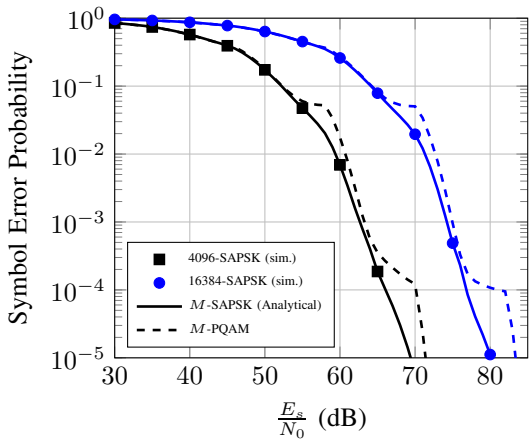

Building on the previous trade-off for the $\Gamma$ value, Figs. \ref{fig:optimal G}a, \ref{fig:optimal G}b, and \ref{fig:optimal G}c illustrate how the optimal value of $\Gamma$ varies with respect to the prevailing noise conditions, the modulation scheme employed, and the order $M$ of the constellation. These figures clearly show that the optimal $\Gamma$ value for SAPSK is always greater than or equal to that of PQAM, which means that SAPSK symbols are arranged in a way that increases robustness to GPN, since an increased $\Gamma$ translates into fewer symbols per energy level. However, this increased GPN robustness should not come at the cost of sacrificing the system's energy efficiency and robustness to AWGN. To this end, using the optimal values of $\Gamma$ for the cases where $M=4096$ and $M=16384$, we show in Figs. \ref{fig:PTQAvsPQAM_opt_G}a and \ref{fig:PTQAvsPQAM_opt_G}b the SEP of SAPSK and PQAM for $\pn^2=10^{-4}$ and $\pn^2=10^{-2}$, respectively, according to the analytical expression given in \eqref{SEP_ptqam_proposed_metric}. As it can be observed, SAPSK has a higher energy efficiency compared to PQAM, offering an SNR gain ranging from 1dB to 3dB over the presented SNR range and for the studied GPN levels. Therefore, Fig. 8 illustrates SAPSK as a more energy-efficient modulation scheme compared to PQAM that also maintains a balance between GPN and AWGN robustness. However, it should be emphasized that although the adaptation of $\Gamma$ based on the prevailing SNR and GPN levels can optimize the performance, it requires an increased system complexity since the transmitter needs to accurately know both the received SNR and GPN levels at the receiver side, thus requiring complex feedback mechanisms to obtain perfect channel state information. Consequently, to meet the system requirements, the used $\Gamma$ value must be predetermined to strike a balance between improved SEP performance and system complexity.

\begin{figure}[h!]
\centering
\begin{minipage}{.5\textwidth}
\centering
\begin{tikzpicture}
	\begin{semilogyaxis}[
    width=0.78\linewidth,
	xlabel = $\frac{E_{s}}{N_{0}}$ (dB),
	ylabel = Symbol Error Probability,
	xmin = 30,
	xmax = 84,
	ymin = 0.00001,
	ymax = 1,
    xtick = {0,10,...,80},
	grid = major,
    legend cell align = {left},
    legend pos = south west,
    legend style={font=\tiny}
	]
\addplot[
color = black,
mark = *,
mark repeat = 1,
mark size = 2,
line width = 1pt,
style = solid
]
table {plots/transaction-data/ptqam/4096-PTQAM_256_-theory-proposed-metric-_1.000000e-04_.txt};
\addlegendentry{SAPSK(4096,256)}

\addplot[
color = blue,
mark = diamond*,
mark repeat = 1,
mark size = 3,
line width = 1pt,
style = solid
]
table {plots/transaction-data/ptqam/4096-PTQAM_1024_-theory-proposed-metric-_1.000000e-04_.txt};
\addlegendentry{SAPSK(4096,1024)}

\addplot[
color = brown,
mark = square*,
mark repeat = 1,
mark size = 2,
line width = 1pt,
style = solid
]
table {plots/transaction-data/ptqam/4096-PTQAM_4096_-_1.000000e-04_.txt};
\addlegendentry{4096-SAPSK(4096)}

\addplot[
color = black,
no marks,
line width = 1pt,
style = dashed,
]
table {plots/transaction-data/pqam/4096-PQAM_256_-theory-approx-v2_1.000000e-04_.txt};
\addlegendentry{$M$-PQAM($\Gamma$)}

\addplot[
color = blue,
no marks,
line width = 1pt,
style = dashed,
]
table {plots/transaction-data/pqam/4096-PQAM_1024_-theory-approx-v2_1.000000e-04_.txt};

\addplot[
color = brown,
no marks,
line width = 1pt,
style = dashed,
]
table {plots/transaction-data/pqam/4096-PQAM_4096_-_1.000000e-04_.txt};
\end{semilogyaxis}
\end{tikzpicture}
\subcaption{$\sigma^2_{\phi} = 10^{-4}$}
\end{minipage}
\begin{minipage}{.5\textwidth}
\centering
\begin{tikzpicture}
\begin{semilogyaxis}[
width=0.78\linewidth,
xlabel = $\frac{E_{s}}{N_{0}}$ (dB),
ylabel = Symbol Error Probability,
xmin = 30,
xmax = 84,
ymin = 0.00001,
ymax = 1,
xtick = {0,10,...,80},
grid = major,
legend cell align = {left},
legend pos = south west,
legend style={font=\tiny}
]
\addplot[
color = black,
mark = *,
mark repeat = 1,
mark size = 2,
line width = 1pt,
style = solid
]
table {plots/transaction-data/ptqam/4096-PTQAM_256_-theory-proposed-metric-_1.000000e-02_.txt};
\addlegendentry{SAPSK(4096,256)}

\addplot[
color = blue,
mark = diamond*,
mark repeat = 1,
mark size = 3,
line width = 1pt,
style = solid
]
table {plots/transaction-data/ptqam/4096-PTQAM_1024_-theory-proposed-metric-_1.000000e-02_.txt};
\addlegendentry{SAPSK(4096,1024)}

\addplot[
color = brown,
mark = square*,
mark repeat = 1,
mark size = 2,
line width = 1pt,
style = solid
]
table {plots/transaction-data/ptqam/4096-PTQAM_4096_-_1.000000e-02_.txt};
\addlegendentry{4096-SAPSK(4096)}

\addplot[
color = black,
no marks,
line width = 1pt,
style = dashed,
]
table {plots/transaction-data/pqam/4096-PQAM_256_-theory-approx-v2_1.000000e-02_.txt};
\addlegendentry{$M$-PQAM($\Gamma$)}

\addplot[
color = blue,
no marks,
line width = 1pt,
style = dashed,
]
table {plots/transaction-data/pqam/4096-PQAM_1024_-theory-approx-v2_1.000000e-02_.txt};

\addplot[
color = brown,
no marks,
line width = 1pt,
style = dashed,
]
table {plots/transaction-data/pqam/4096-PQAM_4096_-_1.000000e-02_.txt};
\end{semilogyaxis}
\end{tikzpicture}
\subcaption{$\sigma^2_{\phi} = 10^{-2}$}
\end{minipage}
\caption{SAPSK vs PQAM for fixed $\Gamma$}
\label{fig:PTQAMvsPQAM_fix_G}
\end{figure}
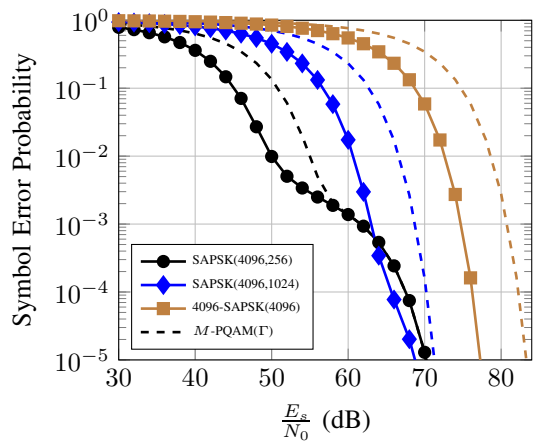
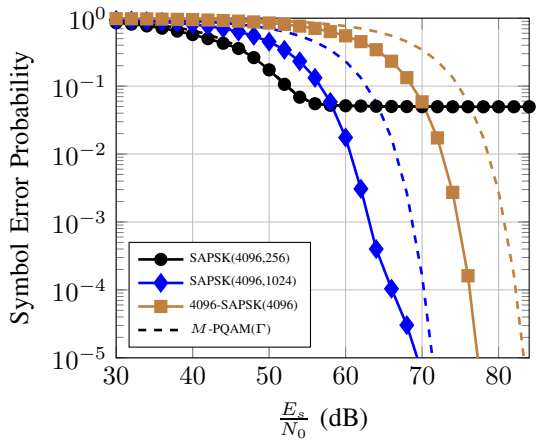

To illustrate the adaptive scheme described for continuously varying the $\Gamma$ of SAPSK to achieve optimal SEP over varying SNRs, Figs. \ref{fig:PTQAMvsPQAM_fix_G}a and \ref{fig:PTQAMvsPQAM_fix_G}b illustrate the SEP performance of an SAPSK and a PQAM constellation with $M=4096$ and $\Gamma$ set to 256, 1024 and 4096, considering GPN levels of $\pn^2=10^{-4}$ and $\pn^2=10^{-2}$. Comparing these variations, we can observe that SAPSK consistently outperforms PQAM in terms of SEP, achieving a gain of 5 dB over PQAM for $\Gamma\in\{1024,2048\}$ for the GPN scenarios considered. In particular, for $\pn^2 = 10^{-4}$, SAPSK($4096$, $256$) shows significant performance gains compared to $4096$-PQAM($256$) up to SNR equal to 55dB, where the SEP reaches $10^{-3}$, while as GPN becomes the dominant distortion, the performance of both constellations is equivalent. In more detail, SAPSK and PQAM have the same angular space between symbols on the same energy level, thus when GPN is dominant compared to AWGN, the performance of SAPSK and PQAM is similar. It should also be noted that the SNR value where GPN is the dominant distortion depends on the value of $\Gamma$ and the prevailing noise condition. Therefore, as shown in Fig.\ref{fig:PTQAMvsPQAM_fix_G}a, for larger values of $\Gamma$, the performance of SAPSK and PQAM has not converged even for SNR values larger than 70dB. Moreover, as shown in Fig. \ref{fig:PTQAMvsPQAM_fix_G}b, the performance of SAPSK($4096$, $256$) and $4096$-PQAM($256$) becomes equivalent for very small values of SNR because the GPN level is increased to $\pn^2=10^{-2}$, reducing the value of SNR where GPN is dominant over AWGN. Finally, it can be observed that in the case where the $4096$-SAPSK constellation is used, with a fixed $\Gamma$ value, $\Gamma = 1024$ can be chosen as a compromise between energy consumption and resilience to the presence of GPN and AWGN.

\begin{figure}[h!]
\centering
\begin{tikzpicture}
\begin{semilogyaxis}[
width=0.78\linewidth,
xlabel = $\frac{E_{s}}{N_{0}}$ (dB),
ylabel = Symbol Error Probability,
xmin = 30,
xmax = 74,
ymin = 0.00001,
ymax = 1,
xtick = {0,10,...,70},
grid = major,
legend cell align = {left},
legend pos = south west,
legend style={font=\tiny}
]

      \addplot[
        color= black,
        only marks,
	mark=*,
	mark repeat = 1,
	mark size = 2,
    ]
	table {plots/transaction-data/ptqam/4096-PTQAM_256_-sim-OPT-metric-_1.000000e-02_.txt};
    \addlegendentry{SAPSK(4096,256) \eqref{optimal_detection_metric}}
    
    \addplot[
        color= blue,
        only marks,
	mark=triangle*,
	mark repeat = 1,
	mark size = 3,
    ]
	table {plots/transaction-data/ptqam/4096-PTQAM_512_-sim-OPT-metric-_1.000000e-02_.txt};
    \addlegendentry{SAPSK(4096,512) \eqref{optimal_detection_metric}}
 
    \addplot[
        color = brown,
        only marks,
	mark=square*,
	mark repeat = 1,
	mark size = 2,
    ]
	table {plots/transaction-data/ptqam/4096-PTQAM_1024_-sim-OPT-metric-_1.000000e-02_.txt};
    \addlegendentry{SAPSK(4096,1024) \eqref{optimal_detection_metric}}
 
    \addplot[
        color= red,
        only marks,
	mark=diamond*,
	mark repeat = 1,
	mark size = 3,
    ]
	table {plots/transaction-data/ptqam/4096-PTQAM_2048_-sim-OPT-metric-_1.000000e-02_.txt};
    \addlegendentry{SAPSK(4096,2048) \eqref{optimal_detection_metric}}
    
    \addplot[
	color = black,
        no marks,
	line width = 1pt,
	style = solid,
    ]
	table {plots/transaction-data/ptqam/4096-PTQAM_256_-sim-OPT-metric-_1.000000e-02_.txt};
    \addlegendentry{Algorithm \ref{optimal_detection_metric}}
    
    \addplot[
	color = blue,
        no marks,
	line width = 1pt,
	style = solid,
    ]
	table {plots/transaction-data/ptqam/4096-PTQAM_512_-sim-OPT-metric-_1.000000e-02_.txt};

    \addplot[
	color = brown,
        no marks,
	line width = 1pt,
	style = solid,
    ]
	table {plots/transaction-data/ptqam/4096-PTQAM_1024_-sim-OPT-metric-_1.000000e-02_.txt};

    \addplot[
	color = red,
        no marks,
	line width = 1pt,
	style = solid,
    ]
	table {plots/transaction-data/ptqam/4096-PTQAM_2048_-sim-OPT-metric-_1.000000e-02_.txt};
\end{semilogyaxis}
\end{tikzpicture}
\caption{Proposed detection algorithm for SAPSK when $\sigma^2_{\phi} = 10^{-2}$}
\label{fig:det_algo}
\end{figure}
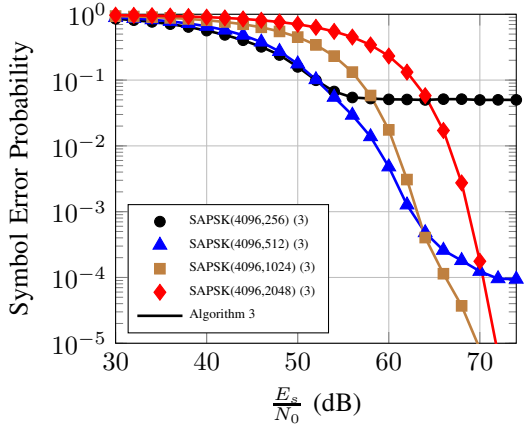

Finally, Fig. \ref{fig:det_algo} illustrates a comparison between the detection efficiency of the proposed detection algorithm for SAPSK constellations, outlined in Algorithm \ref{det_algo}, and the performance achieved by the optimal detector GAP-D, as defined by \eqref{optimal_detection_metric}.  Using GAP-D for symbol-by-symbol detection in the presence of GPN has a complexity of $\mathcal{O}(M)$, since it requires $M$ computations to obtain $\hat{s}$. However, using the decision regions of SAPSK formulated in the polar plane by the proposed GPD-D metric, we have designed Algorithm 1 with a complexity of $\mathcal{O}(1)$. This complexity remains constant, requiring only two computations to obtain $\hat{s}$ regardless of the constellation order $M$, thus making the adaptation of super constellations in communication systems feasible in terms of detection time complexity. 

\section{Conclusion}
In this paper, we introduced a novel modulation scheme, namely SAPSK, that is consistent with the rules of robust super-constellation communications. Specifically, we proposed GPD-D, a novel detector that closely approximates the performance of the optimal maximum likelihood detector GAP-D in the presence of both AWGN and GPN, which describes a weighted Euclidean distance detector in the polar plane. According to our results, we verified the effectiveness of GPD-D, which allows the derivation of a tight, closed-form approximation for the SEP of SAPSK in channels affected by both AWGN and GPN. Furthermore, we demonstrated the superiority of SAPSK over PQAM in terms of energy efficiency in both adaptive and fixed communication links. Finally, we illustrate the detection efficiency of the proposed $\mathcal{O}(1)$ algorithm for SAPSK constellations, paving the way for their practical implementation.

\appendices
\section{Proof of Proposition \ref{prop:orientation}} \label{proof_Prop2}
Initially, to obtain the decision regions of an SAPSK constellation, it is imperative to characterize the scaling that will be induced to the constellation symbols due to the existing noise conditions. In this direction, considering that a symbol $r$ is received such that it satisfies $E_q\leq|r|^2\leq E_{q+1}$ and  using \eqref{dE} and \eqref{approx_first_branch}, the altered vertical and horizontal distances $\delta_{\theta}'$ and $\delta_{\rho}'$ between two neighboring symbols in the polar plane after the application of GPD-D as illustrated in Fig. \ref{fig:PTQAM-decision-regions}, can be expressed as
\begin{equation}\label{delta_rho_prime}
\begin{aligned}
    \delta_{\rho}' = \frac{\delta_{\rho}}{\sigma_{n}}, \\
\end{aligned}
\end{equation}
and
\begin{equation}\label{delta_theta_prime}
\begin{aligned}
    \delta_{\theta}' = \frac{\delta_{\theta}}{\sqrt{\pn^2+\frac{\sigma_{n}^2}{2E_{q}}}}.
\end{aligned}
\end{equation}
For the case where the AWGN is greater than GPN, then the hexagons are oriented along the $\rho$ axis to mitigate the symbol displacements caused by AWGN across consecutive energy levels. However, as GPN increases, the hexagons start to elongate along the $\theta$ axis to compensate for the effect of GPN on the constellation symbols. Furthermore, the elongation of the hexagons along the $\theta$ axis is upper limited by the half distance $\frac{\delta_{\theta}'}{2}$ between two adjacent symbols on the same column due to the NN nature of GPD-D metric. Thus, when the distance between the symbol and the peak of the hexagons along $\theta$ axis equals $\frac{\delta_{\theta}'}{2}$, the orientation $O_h$ changes from $\rho$ to $\theta$ axis undergoing a $90^{\degree}$ rotation. In this direction, we can determine that the hexagonal regions' orientation has changed from $\rho$ axis to $\theta$ axis when the distance $R_d$ which describes the distance between each symbol and the hexagon peak for a hexagon oriented to $\rho$ axis becomes equal to $\frac{\delta_{\theta}'}{2}$. Therefore, according to the decision regions' orientation, the distance $R_{d}$ is given as
\begin{equation}\label{Rd}
\begin{aligned}
  R_{d} =
\begin{cases}
    \begin{split}
       \frac{\sqrt{\delta_{\rho}'^2 + \left(\frac{\delta_{\theta}'}{2}\right)^2}}{2\cos{\left(90^{\degree}-\alpha\right)}},
   \end{split}
      &\textbf{$O_h\equiv\rho$ axis} \\\\
   \begin{split}
       \frac{\sqrt{\delta_{\rho}'^2 + \left(\frac{\delta_{\theta}'}{2}\right)^2}}{2\cos{\alpha}},
   \end{split} 
      &\textbf{$O_h\equiv\theta$ axis}
   \end{cases}
\end{aligned}
\end{equation}
where angle $\alpha$ is depicted in Fig. \ref{fig:PTQAM-decision-regions}. Additionally, by utilizing $\tan{\alpha} = \frac{\delta_{\theta}'}{2\delta_{\rho}'}$ and $\cos\alpha = \frac{1}{\sqrt{1+\tan^2{\alpha}}}$, \eqref{Rd} can be rewritten as\color{black}
\begin{equation}\label{Rtheta_c}
\begin{aligned}
  R_{d} =
\begin{cases}
    \begin{split}
       \frac{\delta_{\theta}'}{2}\left(\frac{\tan^2{\alpha}+1}{2\tan^2{\alpha}}\right),
   \end{split}
      &\textbf{$O_h\equiv\rho$ axis} \\\\
   \begin{split}
       \delta_{\rho}'\left(\frac{\tan^2{\alpha}+1}{2}\right),
   \end{split} 
      &\textbf{$O_h\equiv\theta$ axis}
   \end{cases}
\end{aligned}
\end{equation}
Moreover, by substituting \eqref{dr} in \eqref{delta_rho_prime}, as well as, \eqref{Eq_sxesi} and \eqref{dth} in \eqref{delta_theta_prime}, then it holds that
\begin{equation}\label{tan_alphaSNR}
\begin{aligned}
    \tan{\alpha} = \frac{\pi^2\Gamma^2\left(4\Gamma^2-1\right)}{M^2\sqrt{24\Bar{\gamma}\left(\sigma_{\phi}^2 + \frac{4\Gamma^2-1}{6\Bar{\gamma}(2q-1)^2}\right)}}, 
\end{aligned}
\end{equation}
and by substituting \eqref{tan_alphaSNR} in \eqref{Rtheta_c}, $R_{d}$ can be rewritten as
\begin{equation}\label{Rtheta_c_SNR}
\small
\begin{aligned}
R_d = 
    \begin{cases}
    \begin{split}
       \frac{\delta_{\theta}'}{2} \underbrace{\left(\frac{ 24\Bar{\gamma}\left(\sigma_{\phi}^2 + \frac{4\Gamma^2-1}{6\Bar{\gamma}(2q-1)^2}\right) + \Asquare}{2\Asquare}\right)}_{D_{1,q}},
   \end{split}
      &\textbf{$O_h\equiv\rho$ axis} \\\\
   \begin{split}
       \delta_{\rho}' \underbrace{\left(\frac{24\bar{\gamma}\left(\sigma_{\phi}^2 + \frac{4\Gamma^2-1}{6\Bar{\gamma}(2q-1)^2}\right) + \Asquare}{48\bar{\gamma}\left(\sigma_{\phi}^2 + \frac{4\Gamma^2-1}{6\Bar{\gamma}(2q-1)^2}\right)}\right)}_{D_{2,q}},
   \end{split} 
      &\textbf{$O_h\equiv\theta$ axis}
   \end{cases}
\end{aligned}
\end{equation}
where $A\left(M,\Gamma\right) = \frac{\pi\Gamma\sqrt{4\Gamma^2-1}}{M}$.
Thus, to determine that the hexagonal regions' orientation has changed from $\rho$ axis to $\theta$ axis, i.e., the distance $R_d$ for a hexagon oriented to $\rho$ axis becomes equal to $\frac{\delta_{\theta}'}{2}$, $D_{1,q}$ needs to be equal to 1. To this end, by setting $D_{1,q}=1$, we can obtain the threshold value $\gamma_{q}$ as
\begin{equation}
\begin{aligned}
  \gamma_{q} = \frac{\left(2q-1\right)^2\pi^2\Gamma^2\left(4\Gamma^2-1\right)-4M^2\left(4\Gamma^2-1\right)}{24M^2\left(2q-1\right)^2\sigma_{\phi}^2}, 
\end{aligned}
\end{equation}
which concludes the proof.
\section{Proof of Proposition \ref{prop:PTQAM}} \label{proof_Prop3}
As it can be seen in Fig. \ref{fig:PTQAM-decision-regions}, the hexagonal decision regions can be decomposed by a main rectangular region and two triangles. Thus, for the case where GPD-D is applied and according to the decision regions orientation, the probability of erroneous detection of a symbol $s_{q}$ on the $q$-th energy level of an SAPSK constellation is given as 
\begin{equation}
    P_q = 1 - \left(P_{q,A} + 2 P_{q,B}\right),
\end{equation}
where $P_{q,A}$ is the probability to be within the main rectangle, and $P_{q,B}$ is the probability to lie within a triangle of the hexagonal decision region. To this end, to calculate the SEP for an SAPSK symbol, we need to derive $P_{q,A}$ and $P_{q,B}$.

Initially, the form of the main rectangle and the triangles is affected by the prevailing noise conditions. Therefore, according to the orientation $O_h$ of the decision region and the scaling occurred due to the GPD-D, assuming that $s_{q}$ coincides with the origin $O = (0,0)$ since it is located in the center of the decision region and that $n_{\rho}$ and $n_{\theta}$ are independent RVs, the probability $P_{q, A}$ can be written as 
\begin{equation}\label{PqA}
\small
    \begin{aligned}
        P_{q,A} = 
        \begin{cases}
        \begin{split}
            &P\left( -\delta_{\rho}'\leq \frac{n_{\rho}}{\sqrt{\frac{\sigma_{n}^2}{2}}}\leq \delta_{\rho}'\right)\\
            &\times P\left( -\frac{R_{w}}{2} \leq \frac{n_{\theta}}{\sqrt{\pn^2+\frac{\sigma_{n}^2}{2E_q}}}\leq \frac{R_{w}}{2}\right),
       \end{split}
        &\text{$\bar{\gamma}$} \leq \gamma_q \\\\
       \begin{split}
        &P\left( -\frac{R_{w}}{2} \leq \frac{n_{\rho}}{\sqrt{\frac{\sigma_{n}^2}{2}}}\leq \frac{R_{w}}{2}\right)\\
        &\times P\left( -\frac{\delta_{\theta}'}{2}\leq \frac{n_{\theta}}{\sqrt{\pn^2+\frac{\sigma_{n}^2}{2E_q}}}\leq \frac{\delta_{\theta}'}{2}\right),
       \end{split} 
       &\text{$\bar{\gamma}$} > \gamma_q,
     \end{cases}
    \end{aligned}
\end{equation}
where $R_{w}$ denotes the width of the main rectangle according
to its orientation $O_h$, as shown in Fig. \ref{fig:PTQAM-decision-regions}, and it is given as
\begin{equation} \label{Rw}
    \begin{aligned}
        R_w = 
    \begin{cases}
        \begin{split}
           \delta_{\theta}' - 2R_{d},
       \end{split}
          &\text{$\bar{\gamma}$} \leq \gamma_q \\\\
       \begin{split}
           2\delta_{\rho}' - 2R_{d},
       \end{split} 
          &\text{$\bar{\gamma}$} > \gamma_q
       \end{cases}
    \end{aligned}
\end{equation}
Additionally, by utilizing \eqref{Rtheta_c_SNR}, \eqref{Rw} can be rewritten as 
\begin{equation}
    \begin{aligned}
        R_w = 
    \begin{cases}
        \begin{split}
           \delta_{\theta}'\left(1-D_{1,q} \right),
       \end{split}
          &\text{$\bar{\gamma}$} \leq \gamma_q \\\\
       \begin{split}
           2\delta_{\rho}'\left(1-D_{2,q} \right),
       \end{split} 
          &\text{$\bar{\gamma}$} >\gamma_q
       \end{cases}
    \end{aligned}
\end{equation}
Finally, to calculate \eqref{PqA}, the cumulative distribution function (CDF) of RVs $n_{\rho}$ and $n_{\theta}$ is required.  In this direction, by taking into account  \cite{Eriksson_metrics}, it holds that  $n_{\rho}\sim\mathcal{N}\left(0, \frac{\sigma_{n}^2}{2}\right)$ and $n_{\theta}\sim\mathcal{N}\left(0, \pn^2+\frac{\sigma_{n}^2}{2E_q}\right)$, thus \eqref{PqA} can be rewritten as
\begin{equation}
\small
    \begin{aligned}
        P_{q,A} = 
        \begin{cases}
        \begin{split}
           \left(1-2Q\left(\delta_{\rho}'\right)\right)\left(1-2Q\left(\frac{\delta_{\theta}'}{2}\left(1-D_{1,q}\right)\right)\right),
       \end{split}
        &\text{$\bar{\gamma}$} \leq \gamma_q \\\\
       \begin{split}
          \left(1-2Q\left(\frac{\delta_{\theta}'}{2}\right)\right)\left(1-2Q\left(\delta_{\rho}'\left(1-D_{2,q}\right)\right)\right),
       \end{split} 
       &\text{$\bar{\gamma}$} > \gamma_q,
     \end{cases}
    \end{aligned}
\end{equation}
which concludes the calculation of $P_{q, A}$.

Regarding the calculation of $P_{q, B}$, we can split each triangle into two orthogonal triangles, calculate the probability $P_t$ for the received symbol $r$ to lie within one of these orthogonal triangles, and then multiply it by four. However, the probability of $s_q$ lying inside a triangle does not have a closed-form expression due to the complexity of the bivariate normal distribution and the constraints imposed by the triangle's boundaries. Thus, the area of each triangle can be tightly approximated by $N$ rectangles, as shown in Fig. \ref{fig:PTQAM-decision-regions}, whose dimensions depend on the decision region's orientation. Therefore, considering $O_h$, $P_t$ can be approximated as
\begin{equation}\label{Pt}
\small
    \begin{aligned}
        P_t \approx 
        \begin{cases}
        \begin{split}
            \sum_{z=1}^{N}&P\left( \frac{(z-1)\delta_{\rho}'}{N}\leq \frac{n_{\rho}}{\sqrt{\frac{\sigma_{n}^2}{2}}}\leq \frac{z\delta_{\rho}'}{N}\right)\\
            &\times P\left( \frac{R_{w}}{2} \leq \frac{n_{\theta}}{\sqrt{\pn^2+\frac{\sigma_{n}^2}{2E_q}}}\leq \frac{B_{z,1}\delta_{\theta}}{2}\right),
       \end{split}
        &\text{$\bar{\gamma}$} \leq \gamma_q \\\\
       \begin{split}
        \sum_{z=1}^{N}&P\left( \frac{\left(z-1\right)\delta_{\theta}'}{2N} \leq \frac{n_{\rho}}{\sqrt{\frac{\sigma_{n}^2}{2}}}\leq \frac{z\delta_{\theta}'}{2N}\right)\\
        &\times P\left(\frac{R_{w}}{2}\leq \frac{n_{\theta}}{\sqrt{\pn^2+\frac{\sigma_{n}^2}{2E_q}}}\leq B_{z,2}\delta_{\rho}'\right),
       \end{split} 
       &\text{$\bar{\gamma}$} > \gamma_q,
     \end{cases}
    \end{aligned}
\end{equation}
where $B_{z,1} = \frac{\left(1-2D_{1,q}\right)z}{N} + D_{1,q}$, and $B_{z,2} = \frac{\left(1-2D_{2,q}\right)z}{N} + D_{2,q}$. Finally, by leveraging the CDFs of $n_{\rho}$ and $n_{\theta}$, $P_t$ can be derived, thus $P_{q, B}$ can be calculated as $P_{q,B}=4 P_t$, which concludes the proof.

\bibliographystyle{IEEEtran}
\bibliography{Bibliography}

\begin{thebibliography}{10}
\providecommand{\url}[1]{#1}
\csname url@samestyle\endcsname
\providecommand{\newblock}{\relax}
\providecommand{\bibinfo}[2]{#2}
\providecommand{\BIBentrySTDinterwordspacing}{\spaceskip=0pt\relax}
\providecommand{\BIBentryALTinterwordstretchfactor}{4}
\providecommand{\BIBentryALTinterwordspacing}{\spaceskip=\fontdimen2\font plus
\BIBentryALTinterwordstretchfactor\fontdimen3\font minus
  \fontdimen4\font\relax}
\providecommand{\BIBforeignlanguage}[2]{{%
\expandafter\ifx\csname l@#1\endcsname\relax
\typeout{** WARNING: IEEEtran.bst: No hyphenation pattern has been}%
\typeout{** loaded for the language `#1'. Using the pattern for}%
\typeout{** the default language instead.}%
\else
\language=\csname l@#1\endcsname
\fi
#2}}
\providecommand{\BIBdecl}{\relax}
\BIBdecl

\bibitem{6G}
Z.~Zhang, Y.~Xiao, Z.~Ma, M.~Xiao, Z.~Ding, X.~Lei, G.~K. Karagiannidis, and
  P.~Fan, ``6{G} wireless networks: Vision, requirements, architecture, and key
  technologies,'' \emph{IEEE Veh. Technol. Mag.}, vol.~14, no.~3, pp. 28--41,
  2019.

\bibitem{liaskos}
{C. Liaskos} \emph{et~al.}, ``{XR-RF} imaging enabled by software-defined
  metasurfaces and machine learning: Foundational vision, technologies and
  challenges,'' \emph{IEEE Access}, vol.~10, pp. 119\,841--119\,862, 2022.

\bibitem{mmWave}
M.~Xiao, S.~Mumtaz, Y.~Huang, L.~Dai, Y.~Li, M.~Matthaiou, G.~Karagiannidis,
  E.~Björnson, K.~Yang, C.-L. I., and A.~Ghosh, ``Millimeter wave
  communications for future mobile networks,'' \emph{IEEE J. Sel. Areas
  Commun.}, vol.~pp, 05 2017.

\bibitem{massive_mimo_karag}
S.~K. Goudos, P.~D. Diamantoulakis, and G.~K. Karagiannidis, ``Multi-objective
  optimization in {5G} wireless networks with massive {MIMO},'' \emph{IEEE
  Commun. Lett.}, vol.~22, no.~11, pp. 2346--2349, 2018.

\bibitem{1k-QAM_fiber_optics}
M.~P. Yankov, D.~Zibar, K.~J. Larsen, L.~P.~B. Christensen, and S.~Forchhammer,
  ``Constellation shaping for fiber-optic channels with {QAM} and high spectral
  efficiency,'' \emph{IEEE Photon. Technol. Lett.}, vol.~26, no.~23, pp.
  2407--2410, 2014.

\bibitem{1k-QAM-ofdm}
Y.~Liu, Y.~Wang, and B.~Ai, ``An efficient ace scheme for {PAPR} reduction of
  {OFDM} signals with high-order constellation,'' \emph{IEEE Access}, vol.~7,
  pp. 118\,322--118\,332, 2019.

\bibitem{1k-QAM-low-complexity-demap}
M.~Fuentes, D.~Vargas, and D.~Gómez-Barquero, ``Low-complexity demapping
  algorithm for two-dimensional non-uniform constellations,'' \emph{IEEE Trans.
  Broadcast.}, vol.~62, no.~2, pp. 375--383, 2016.

\bibitem{1k-QAM_uhdtv}
B.~Mouhouche, D.~Ansorregui, and A.~Mourad, ``High order non-uniform
  constellations for broadcasting {UHDTV},'' in \emph{Proc. IEEE Wireless
  Communications and Networking Conference (WCNC)}, 2014, pp. 600--605.

\bibitem{1k-QAM_bicm}
M.~Yankov, S.~Forchhammer, K.~J. Larsen, and L.~P.~B. Christensen,
  ``Rate-adaptive constellation shaping for near-capacity achieving turbo coded
  {BICM},'' in \emph{Proc. IEEE International Conference on Communications
  (ICC)}, 2014, pp. 2112--2117.

\bibitem{1k-non_uniform_QAM}
J.~Zoellner and N.~Loghin, ``Optimization of high-order non-uniform {QAM}
  constellations,'' in \emph{Proc. IEEE International Symposium on Broadband
  Multimedia Systems and Broadcasting (BMSB)}, 2013, pp. 1--6.

\bibitem{MIMO-256QAM}
J.~Céspedes, P.~M. Olmos, M.~Sánchez-Fernández, and F.~Perez-Cruz,
  ``Expectation propagation detection for high-order high-dimensional {MIMO}
  systems,'' \emph{IEEE Trans. Commun.}, vol.~62, no.~8, pp. 2840--2849, 2014.

\bibitem{MIMO-equalizer-high-orders}
Q.~Wang, ``Nested variational chain and its application in massive {MIMO}
  detection for high-order constellations,'' \emph{Entropy}, vol.~25, no.~12,
  2023.

\bibitem{MIMO-PSK}
B.~Sun, Y.~Zhou, J.~Yuan, Y.-F. Liu, L.~Wang, and L.~Liu, ``High order psk
  modulation in massive {MIMO} systems with 1-bit adcs,'' \emph{IEEE Trans.
  Wireless Commun.}, vol.~20, no.~4, pp. 2652--2669, 2021.

\bibitem{mmWave-256QAM}
I.~Kallfass, J.~Antes, T.~Schneider, F.~Kurz, D.~Lopez-Diaz, S.~Diebold,
  H.~Massler, A.~Leuther, and A.~Tessmann, ``All active {MMIC}-based wireless
  communication at 220 {GH}z,'' \emph{IEEE Trans. Terahertz Sci. Technol.},
  vol.~1, no.~2, pp. 477--487, 2011.

\bibitem{4k-QAM-ofdm}
L.~zhao, K.~Wang, and W.~Zhou, ``Transmission of 4096-{QAM} {OFDM} at
  {D}-band,'' \emph{Opt. Express}, vol.~31, no.~2, pp. 2270--2281, 2023.

\bibitem{1k-OFDM-28Ghz}
L.~Zhao, B.~Sang, Y.~Tan, K.~Wang, W.~Zhou, Y.~Wang, J.~Ding, J.~Xiao, and
  J.~Yu, ``Transmission of 1024-{QAM} {OFDM} at 28 {GH}z radio frequency using
  5{G} millimeter wave phased array antenna,'' \emph{IEEE Trans. Microw. Theory
  Tech.}, vol.~70, no.~9, pp. 4211--4217, 2022.

\bibitem{hqam_th}
T.~K. Oikonomou, S.~A. Tegos, D.~Tyrovolas, P.~D. Diamantoulakis, and G.~K.
  Karagiannidis, ``On the error analysis of hexagonal-{QAM} constellations,''
  \emph{IEEE Commun. Lett.}, vol.~26, no.~8, pp. 1764--1768, 2022.

\bibitem{FSO-high-order}
P.~K. Singya, N.~Kumar, V.~Bhatia, and M.-S. Alouini, ``On the performance
  analysis of higher order {QAM} schemes over mixed {RF/FSO} systems,''
  \emph{IEEE Trans. Veh. Technol.}, vol.~69, no.~7, pp. 7366--7378, 2020.

\bibitem{HQAM_ris}
T.~K. Oikonomou, D.~Tyrovolas, S.~A. Tegos, P.~D. Diamantoulakis,
  P.~Sarigiannidis, C.~Liaskos, and G.~K. Karagiannidis, ``On the performance
  of {RIS}-assisted networks with {HQAM},'' 2024.

\bibitem{LO}
A.~Hajimiri and T.~Lee, ``A general theory of phase noise in electrical
  oscillators,'' \emph{IEEE J. Solid-State Circuits}, vol.~33, no.~2, pp.
  179--194, 1998.

\bibitem{AMC-pn}
T.~K. Oikonomou, N.~G. Evgenidis, D.~G. Nixarlidis, D.~Tyrovolas, S.~A. Tegos,
  P.~D. Diamantoulakis, P.~G. Sarigiannidis, and G.~K. Karagiannidis,
  ``{CNN}-based automatic modulation classification under phase
  imperfections,'' \emph{IEEE Wireless Commun. Lett.}, vol.~13, no.~5, pp.
  1508--1512, 2024.

\bibitem{1kand4k-QAM-ofdm-pn}
H.~Otsuka, R.~Tian, and K.~Senda, ``Transmission performance of an {OFDM}-based
  higher-order modulation scheme in multipath fading channels,'' \emph{Journal
  of Sensor and Actuator Networks}, vol.~8, no.~2, 2019.

\bibitem{pn_qam}
A.~V. Menon, A.~Gunjegai, Aishwarya, and D.~G. Kurup, ``Combined amplitude and
  phase noise effects in {QAM} direct conversion receivers,'' in \emph{Proc.
  International Conference on Microwave, Optical and Communication Engineering
  (ICMOCE)}, 2015, pp. 346--348.

\bibitem{Eriksson_opt}
R.~Krishnan, A.~Graell~i Amat, T.~Eriksson, and G.~Colavolpe, ``Constellation
  optimization in the presence of strong phase noise,'' \emph{IEEE Trans.
  Commun.}, vol.~61, no.~12, pp. 5056--5066, 2013.

\bibitem{spiral_const}
A.~Ugolini, A.~Piemontese, and T.~Eriksson, ``Spiral constellations for phase
  noise channels,'' \emph{IEEE Trans. on Commun.}, vol.~67, no.~11, pp.
  7799--7810, 2019.

\bibitem{bicais_pqam}
S.~Bicaïs and J.-B. Doré, ``Design of digital communications for strong phase
  noise channels,'' \emph{IEEE Open J. Veh. Technol.}, vol.~1, pp. 227--243,
  2020.

\bibitem{proakis2002communication}
J.~G. Proakis and M.~Salehi, \emph{Communication Systems Engineering}.\hskip
  1em plus 0.5em minus 0.4em\relax Prentice Hall, 2002.

\bibitem{akyldiz}
E.~Khorov, A.~Kureev, I.~Levitsky, and I.~F. Akyildiz, ``A phase noise
  resistant constellation rotation method and its experimental validation for
  {NOMA} {W}i-{F}i,'' \emph{IEEE J. Sel. Areas Commun.}, vol.~40, no.~4, pp.
  1346--1354, 2022.

\bibitem{bicais_pn_levels}
S.~Bicaïs, J.-B. Doré, G.~Gougéon, and Y.~Corre, ``Optimized single carrier
  transceiver for future sub-terahertz applications,'' in \emph{Proc. IEEE
  International Conference on Acoustics, Speech and Signal Processing
  (ICASSP)}, 2020, pp. 5095--5099.

\bibitem{Eriksson_metrics}
R.~Krishnan, M.~R. Khanzadi, T.~Eriksson, and T.~Svensson, ``Soft metrics and
  their performance analysis for optimal data detection in the presence of
  strong oscillator phase noise,'' \emph{IEEE Trans. Commun.}, vol.~61, 10
  2013.

\end{thebibliography}

\end{document}